\documentclass{emulateapj}

\usepackage{graphicx}
\usepackage{psfig}
\usepackage{epsfig}
\usepackage{url}
\usepackage{hyperref}
\usepackage{breakurl}

\def\amin{\ifmmode^{\prime}\else$^{\prime}$\fi}
\def\asec{\ifmmode^{\prime\prime}\else$^{\prime\prime}$\fi}
\def\aap{A\&A}

\def\apj{ApJ}
\def\apjs{ApJS}
\def\apjl{ApJL}

\def\mnras{MNRAS}
\def\aj{AJ}
\def\nat{Natur}
\def\aaps{A\&A Supp.}

\def\pasp{PASP}

\def\araa{ARA\&A}

\def\pasa{PASA}

\shorttitle{M59-UCD3}
\shortauthors{Ahn et al.}

\begin{document}

%\slugcomment{Resubmitted Version, Feb. 26, 2010}

\title{The Black Hole in the Most Massive Ultracompact Dwarf Galaxy M59-UCD3}

\author{Christopher P. Ahn\altaffilmark{1,2}, Anil C. Seth\altaffilmark{1}, Michele Cappellari\altaffilmark{3}, Davor Krajnovi\'c\altaffilmark{4}, Jay Strader\altaffilmark{5}, Karina T. Voggel\altaffilmark{1}, Jonelle L. Walsh\altaffilmark{6}, Arash Bahramian\altaffilmark{5}, Holger Baumgardt\altaffilmark{7}, Jean Brodie\altaffilmark{8,9}, Igor Chilingarian\altaffilmark{10,11}, Laura Chomiuk\altaffilmark{5}, Mark den Brok\altaffilmark{4}, Matthias Frank\altaffilmark{12}, Michael Hilker\altaffilmark{13}, Richard M. McDermid\altaffilmark{14,15,16}, Steffen Mieske\altaffilmark{17}, Nadine Neumayer\altaffilmark{18}, Dieu D. Nguyen\altaffilmark{1}, Renuka Pechetti\altaffilmark{1}, Aaron J. Romanowsky\altaffilmark{8,19}, Lee Spitler\altaffilmark{14,15,16}}

\altaffiltext{1}{University of Utah. Department of Physics \& Astronomy, 115 S 1400 E, Salt Lake City, UT 84105}
\altaffiltext{2}{{\tt chris.ahn43@gmail.com}}
\altaffiltext{3}{Sub-Department of Astrophysics, Department of Physics, University of Oxford, Denys Wilkinson Building, Keble Road, Oxford, OX1 3RH, UK}
\altaffiltext{4}{Leibniz-Institut f\"ur Astrophysik Potsdam (AIP), An der Sternwarte 16, D-14482 Potsdam, Germany}
\altaffiltext{5}{Michigan State University}
\altaffiltext{6}{George P. and Cynthia Woods Mitchell Institute for Fundamental Physics and Astronomy, Department of Physics and Astronomy, Texas A\&M University, College Station, TX 77843}
\altaffiltext{7}{University of Queensland}
\altaffiltext{8}{University of California Observatories, 1156 High Street, Santa Cruz, CA 95064, USA}
\altaffiltext{9}{University of California Santa Cruz}
\altaffiltext{10}{Smithsonian Astrophysical Observatory, 60 Garden St. MS09, 02138 Cambridge, MA, USA}
\altaffiltext{11}{Sternberg Astronomical Institute, M.V. Lomonosov Moscow State University, 13 Universitetsky prospect, 119992 Moscow, Russia}
%\altaffiltext{12}{ETH Zurich, Switzerland}
\altaffiltext{12}{Landessternwarte, Zentrum f\"ur Astronomie der Universit\"at Heidelberg, K\"onigsstuhl 12, D-69117 Heidelberg, Germany}
\altaffiltext{13}{European Southern Observatory, Garching}
\altaffiltext{14}{Australian Astronomical Observatory, PO Box 915 North Ryde NSW 1670, Australia}
\altaffiltext{15}{Research Centre for Astronomy, Astrophysics \& Astrophotonics, Macquarie University, Sydney, NSW 2109, Australia}
\altaffiltext{16}{Department of Physics \& Astronomy, Macquarie University, Sydney, NSW 2109, Australia}
\altaffiltext{17}{European Southern Observatory Alonso de Cordova 3107, Vitacura, Chile}
\altaffiltext{18}{Max-Planck-Institut f\"ur Astronomie}
\altaffiltext{19}{Department of Physics \& Astronomy, San Jos\'e State University, One Washington Square, San Jose, CA 95192, USA}

%\altaffiltext{3}{ETH Zurich, Switzerland}
%\altaffiltext{4}{Michigan State University}
%\altaffiltext{5}{University of Queensland}
%\altaffiltext{6}{Max-Planck-Institut f\"ur Astronomie}
%\altaffiltext{7}{Smithsonian Astrophysical Observatory, 60 Garden St. MS09, 02138 Cambridge, MA, USA}
%\altaffiltext{8}{Sternberg Astronomical Institute, M.V. Lomonosov Moscow State University, 13 Universitetsky prospect, 119992 Moscow, Russia}
%\altaffiltext{9}{Landessternwarte, Zentrum f\"ur Astronomie der Universit\"at Heidelberg, K\"onigsstuhl 12, D-69117 Heidelberg, Germany}
%\altaffiltext{10}{European Southern Observatory, Garching}
%\altaffiltext{11}{Australian Astronomical Observatory, PO Box 915 North Ryde NSW 1670, Australia}
%\altaffiltext{12}{San Jose State University}
%\altaffiltext{13}{University of California Observatories/Lick Observatory}
%\altaffiltext{14}{Research Centre for Astronomy, Astrophysics \& Astrophotonics, Macquarie University, Sydney, NSW 2109, Australia}
%\altaffiltext{15}{Department of Physics \& Astronomy, Macquarie University, Sydney, NSW 2109, Australia}
%\altaffiltext{16}{University of California Santa Cruz}
%\altaffiltext{17}{George P. and Cynthia Woods Mitchell Institute for Fundamental Physics and Astronomy, Department of Physics and Astronomy, Texas A\&M University, College Station, TX 77843}

\begin{abstract}
  We examine the internal properties of the most massive ultracompact dwarf galaxy (UCD), M59-UCD3, by combining adaptive optics assisted near-IR integral field spectroscopy from Gemini/NIFS, and {\it Hubble Space Telescope (HST)} imaging. We use the multi-band {\it HST} imaging to create a mass model that suggests and accounts for the presence of multiple stellar populations and structural components. We combine these mass models with kinematics measurements from Gemini/NIFS to find a best-fit stellar mass-to-light ratio ($M/L$) and black hole (BH) mass using Jeans Anisotropic Models (JAM), axisymmetric Schwarzschild models, and triaxial Schwarzschild models. The best fit parameters in the JAM and axisymmetric Schwarzschild models have black holes between 2.5 and 5.9 million solar masses. The triaxial Schwarzschild models point toward a similar BH mass, but show a minimum $\chi^2$ at a BH mass of $\sim 0$. Models with a BH in all three techniques provide better fits to the central $V_{rms}$ profiles, and thus we estimate the BH mass to be $4.2^{+2.1}_{-1.7} \times 10^{6}$~M$_\odot$ (estimated 1$\sigma$ uncertainties).  We also present deep radio imaging of M59-UCD3 and two other UCDs in Virgo with dynamical BH mass measurements, and compare these to X-ray measurements to check for consistency with the fundamental plane of BH accretion. We detect faint radio emission in M59cO, but find only upper limits for M60-UCD1 and M59-UCD3 despite X-ray detections in both these sources. The BH mass and nuclear light profile of M59-UCD3 suggests it is the tidally stripped remnant of a $\sim$10$^{9-10}$~M$_\odot$ galaxy.  

  %We test for the presence of a central massive black hole (BH) using three separate dynamical modeling techniques.
  %The kinematic data are binned together using the Voronoi binning method and fitted using the penalized pixel fitting (pPXF) algorithm. In each bin, we measured the line-of-sight velocity distribution described by the radial velocity, velocity dispersion, skewness, and kurtosis.
    %but formally also indicate a secondary and apparently unphysical $\chi^2$ minimum with a BH mass $\sim 0$.
  %  We detect a central massive BH in two of the three dynamical modeling techniques with $M_{BH} = 5.9 \pm 3.1 \times 10^6 M_\odot$, $M_{BH} = 2.5^{+7.4}_{-2.4} \times 10^6$ for the JAM and axisymmetric Schwarzschild models, respectively (3$\sigma$ uncertainties). The triaxial Schwarzschild models are consistent with no BH. However, the best-fit triaxial Schwarzschild model does not provide a good fit to the kinematic maps. For this reason, we prefer the JAM and axisymmetric Schwarzschild models.
\end{abstract}


\section{Introduction} \label{sec:intro}
The past few decades have seen the emergence of compact stellar systems that have blurred the conventional lines, based on properties such as mass, luminosity and size, between star clusters and galaxies \citep{hilker99,drinkwater00}. One such classification of objects, the ultracompact dwarf galaxies (UCDs), occupy the region between globular clusters (GCs) and compact ellipticals (cEs) \citep[e.g.][]{misgeldhilker11,brodie11,pfeffer13,norris14,janz16}. With masses and radii of $M >2 \times 10^6$ $M_\odot$ and $r > 10$ pc, UCDs are among the densest stellar systems in the Universe. However, the nature and origin of these dense objects is still widely debated. Early interpretations suggested that UCDs could be the most massive GCs \citep[e.g.][]{mieske02,fellhauer02,fellhauer05,kissler06,murray09} or possibly the tidally stripped remnants of dwarf galaxies \citep[]{bekki01,bekki03,drinkwater03,strader13,pfeffer13,forbes14}. However, there is evidence that both formation mechanisms could contribute to the UCD population we observe \citep{darocha11,brodie11,norriskap11,janz16}. 

In the last decade, observational results based on structural information from {\it Hubble Space Telescope (HST)} imaging combined with integrated velocity dispersion measurements have shown an interesting trend: UCDs dynamical mass-to-light ratios ($M/L_{dyn}$) appear to be systematically elevated when compared to the canonical stellar population estimates \citep{hasegan05,mieske08,dabringhausen08,taylorm10,frank11,strader13}. These results prompted suggestions of variations in the stellar initial mass function (IMF) of UCDs (top-heavy: \citealt{murray09,dabringhausen09,dabringhausen10}; bottom-heavy: \citealt{mieskekroupa08}). Further explanations have suggested that these elevated $M/L$s could be explained by ongoing tidal stripping \citep{forbes14,janz16}, or, as a relic of a massive progenitor galaxy in the tidal stripping scenario, a central massive black hole (BH) making up $\sim10-15$\% of the total mass \citep{mieske13}. Supermassive black holes (SMBHs) have been confirmed in four UCDs with masses $M > 10^7$ $M_\odot$; three in the Virgo cluster \citep{seth14,ahn17}, and one in the Fornax cluster  \citep{afanasiev18}. A search for SMBHs in two lower mass ($M < 10^7$ $M_\odot$) UCDs in Centaurus A yielded a non-detection \citep{voggel18}. However, \citet{voggel18} also showed that the dynamical-to-stellar $M/L$s were overestimated in previous studies. The combination of this evidence still supports the idea that most UCDs with apparently high dynamical-to-stellar mass ratios (including a vast majority of UCDs above $10^7$~M$_\odot$) host SMBHs. Lower mass UCDs may be the high mass end of the globular cluster distribution. This view would be consistent with the analysis of the stripped nuclei contribution to UCDs in $\Lambda$CDM simulations by \citet{pfeffer14} and \citet{pfeffer16}. 

Despite the fact that all of the detected SMBHs are found in massive (M $> 10^7$~M$_\odot$) UCDs, the most massive UCD discovered to date, M59-UCD3 ($M_* \sim 2\times 10^8$ $M_\odot$ $r_e \sim 25$ pc), has been left out. This is in part due to its recent discovery \citep{sandoval15,liu15}, and the lack of high resolution imaging data needed for dynamical modeling. Thus, M59-UCD3 serves as an important test of the idea that the most massive UCDs host SMBHs. In this paper we present the dynamical modeling techniques and results for M59-UCD3.

\begin{figure}[h!]
  \centering
  \includegraphics[trim={0 0 0cm 0cm},clip,scale=0.37]{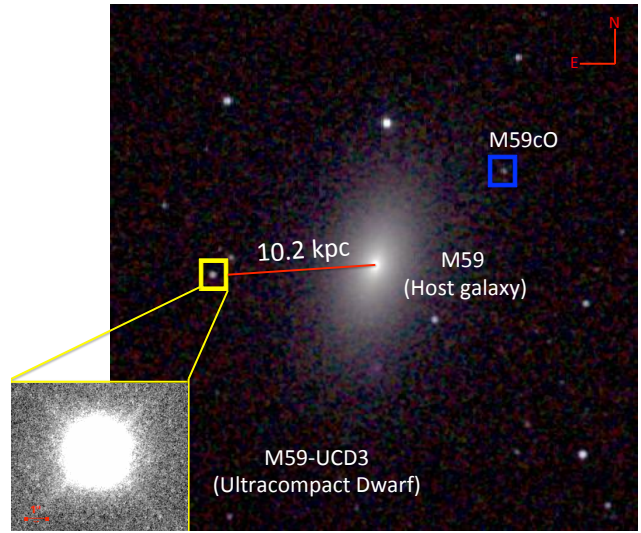}
  \caption{The M59/M59-UCD3 system discussed in this paper. Here, the main image shows the $2MASS$ $LGA$ image \citep{jarrett03}. M59-UCD3 is outlined in the yellow box. The inset image is a zoom-in {\it HST} image taken through the F814W filter on the WFC3 instrument. We also outline another UCD in this system, M59cO, in blue. The red line connecting the UCD to the host galaxy shows the projected distance assuming M59-UCD3 is at a distance of 16.5 Mpc.}
  \label{fig:image}
\end{figure}

An image of M59-UCD3 and its host galaxy (M59$=$NGC~4621) is shown in Figure~\ref{fig:image}. M59-UCD3 is located 10.2 kpc in projection from the center of M59, assuming an average distance of 16.5 Mpc to the Virgo Cluster based on surface brightness fluctations \citep{mei07}. We note that the individual distance of M59 has been measured to be $14.9 \pm 0.4$ Mpc \citep{mei07}, and this distance has been used in previous luminosity and mass estimates; our assumed 16.5~Mpc will yield a 10\% higher dynamical mass estimate relative to the previous mass determination \citep{liu15}, while at 16.5 Mpc M59-UCD3 has a measured $M_V = -14.8$ \citep{sandoval15}. We adopt the conventional definition of $\Gamma\equiv (M/L)_{dyn} / (M/L)_{*}$, which is the ratio between the dynamically determined total $M/L$ and the stellar $M/L$ inferred from stellar population modeling. Throughout this paper, we assume a Chabrier IMF for the stellar population models. The metallicity of M59-UCD3 has been estimated to be near solar, with [Fe/H]$\sim$ $-0.01$ and [$\alpha$/Fe]$\sim$0.21 \citep{sandoval15,liu15,janz16,villaume17}. These values of near solar metallicity and moderate alpha-element enhancement are consistent with previously measured high mass UCDs \citep{evstigneeva07,chilingarianmamon08,firth09,francis12,strader13,janz16}. All magnitudes are reported in the AB magnitude system. Furthermore, all magnitudes and colors have been extinction corrected using $A_{F475W} = 0.100$ and $A_{F814W} = 0.052$ \citep{schlafly11}.

This paper is organized as follows: Section~\ref{sec:data} discusses the data used for analysis, how we determined a density profile, and how the kinematics were modeled. In Section~\ref{sec:model} we present our three dynamical modeling techniques and the results from each. Section~\ref{sec:xrayradio} discusses the radio/X-ray observations of UCDs and whether these observations can be used to infer the presence or not of an accreting SMBH. In Section~\ref{sec:discussconclude} we discuss the implications of the results and present our conclusions. 

\section{Data and Methods} \label{sec:data}

In this section we present the data and our reduction techniques. Section~\ref{hstphoto} discusses the {\it HST} images and our methods for deriving a mass model. Section~\ref{gemspec} explains the reduction of our Gemini/NIFS integral field spectroscopy and the derivation of the kinematics.

\subsection{Imaging data and Deriving a Mass Model} \label{hstphoto}

We obtained images of M59-UCD3 from the {\it HST} GO Cycle 23 program 14067 (PI: Ahn) with the Wide Field Camera 3 (WFC3) instrument, which has a pixel scale of 0.04$\asec$ pixel$^{-1}$. Our data were taken through the F475W and F814W filters. The exposure times in each filter were 1470s and 747s for F475W and F814W, respectively.

We derived a point spread function (PSF) for each filter following the procedure outlined in previous studies \citep{evstigneeva07,ahn17}. To briefly summarize, we generated the distorted PSF with TinyTim and placed these PSFs in an empty copy of the raw {\it HST} flat fielded image at the location of our observed target. The distorted PSFs were then passed through MultiDrizzle using the same parameters as were used for the data. This produces model PSFs that are processed in the same way as the original data.

\begin{figure}[h!]
  \centering
  \includegraphics[trim={0 0 4.3cm 10cm},clip,scale=0.5]{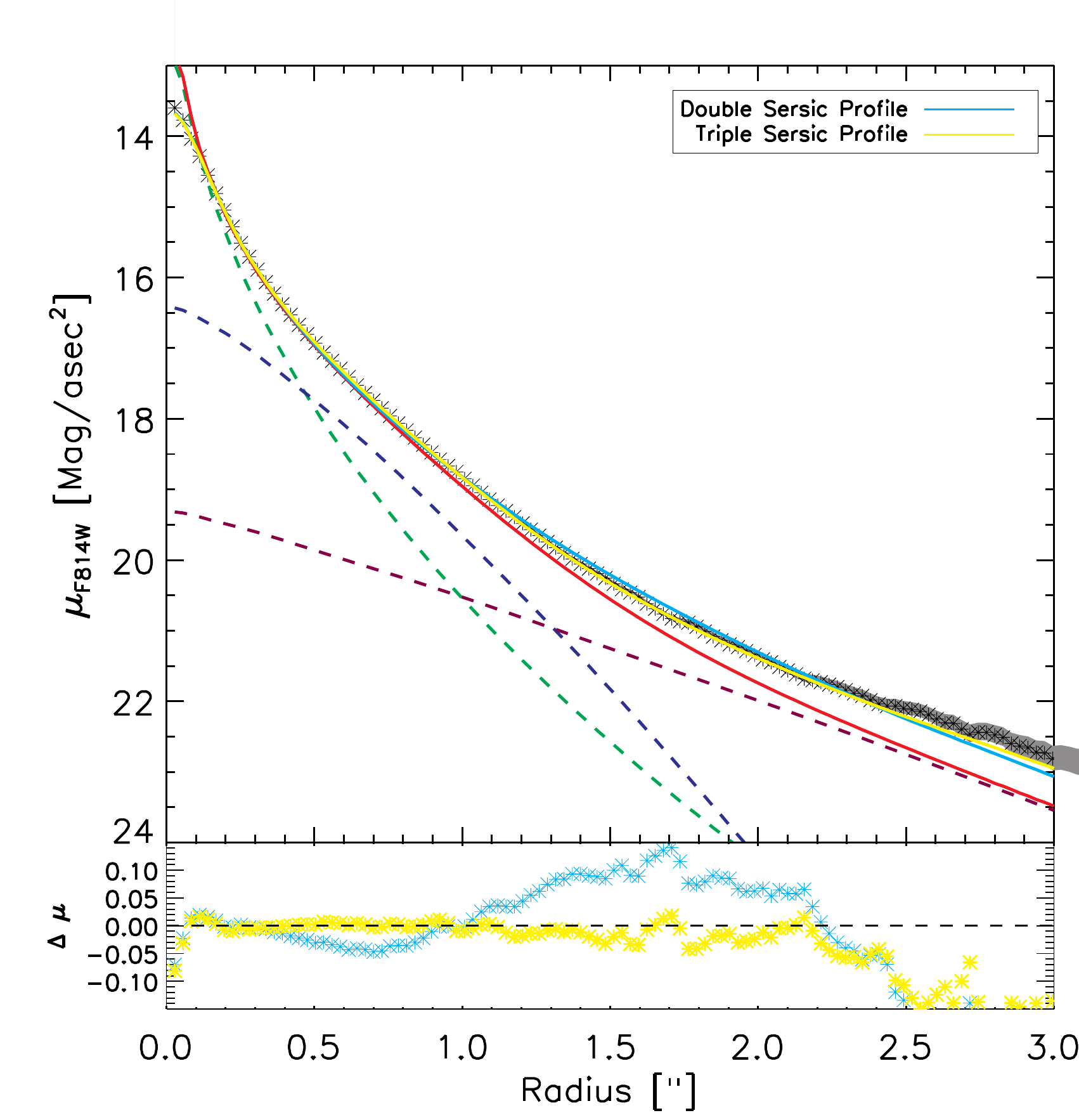}
  \caption{Surface brightness profile of M59-UCD3 in {\it HST}/F814W, which was used for dynamical modeling. Black stars are data, cyan lines are convolved double S\'ersic profile models, yellow lines are convolved triple S\'ersic profile models, the red line is the triple S\'ersic reconstructed profile, and green, blue, and purple lines are the individual S\'ersic components. The gray bands represent the uncertainty in our background sky determination. The residuals between the data and convolved models are shown in the bottom panel.}
  \label{fig:sb}
\end{figure}

\begin{figure}[h!]
  \centering
  \includegraphics[trim={3.1cm 13cm 1cm 1cm},clip,scale=0.55]{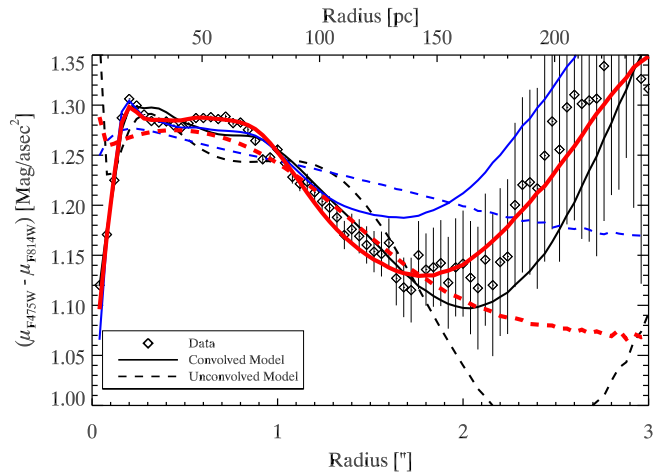}
  \caption{Color profile of M59-UCD3 shown as black diamonds. The error bars are calculated from the uncertainty in our background (sky) level determinations. The solid lines indicate the triple-component S\'ersic model fits that have been convolved with the {\it HST} PSF. Dashed lines show models that are unconvolved. The colors represent whether the shape parameters of the S\'ersic profiles were independent (black) or fixed (red and blue). Blue lines indicate that the shape parameters of the F475W filter were held fixed to the best-fit F814W S\'ersic models, and red lines are vice-versa. The unconvolved fixed models (red and blue) provide a well defined color for each S\'ersic component. Our default model is shown in red. Here, the inner, middle, and outer colors are 1.26, 1.32, and 1.06 mag for our default model, respectively. See Section~\ref{hstphoto} for a discussion on our choice of the default model.}
  \label{fig:color}
\end{figure}

The background (sky) level is traditionally determined from empty portions of the image. However, UCDs generally fall within the stellar halo of their host galaxy. Therefore, the sky level is not uniform across the image. To account for this, we added the MultiDrizzle level, subtracted by the {\it HST} reduction pipeline, back in and modeled the sky as a tilted plane. This was accomplished by masking all foreground/background objects including our UCD in the image. The good pixels (determined from the DQ extension of the image) were then weighted by their corresponding errors. Finally, a plane was fitted to the image to represent the sky level. The formal uncertainties on the sky level determination in this method are negligible. However, a clear systematic effect is seen in that the mean value of the data minus sky model is offset from zero. We regard this as indicative of the systematic uncertainty, which are 0.86 counts in F475W and 1.38 counts in F814W. We use these uncertainties for plotting purposes only in the surface brightness profile (Figure~\ref{fig:sb} as grey bands), and color profile (Figure~\ref{fig:color} as our error bars on the data).

To enable dynamical modeling of M59-UCD3, we needed to create a model to represent the luminosity and mass distribution. Typically, in compact objects such as UCDs, the mass is assumed to trace the light \citep[e.g.][]{mieske13,seth14}. However, previous studies have found significant color gradients in UCDs, suggesting multiple stellar populations \citep{chilingarianmamon08,evstigneeva08,ahn17}. Therefore, two filter data is essential for determining the most accurate luminosity and mass profiles of UCDs. The uncertainties in the luminosity and mass profile combinations are discussed in Section~\ref{sec:JAM}. For now, we discuss the general procedure for determining our luminosity and mass distributions.

The surface brightness profile was determined by fitting the data in each filter to a PSF-convolved, multiple component S\'ersic profile using the two-dimensional fitting algorithm, GALFIT \citep{peng02}. The parameters of the individual S\'ersic profiles that were fitted include: the total magnitude ($m_{tot}$), effective radius ($R_e$), S\'ersic exponent ($n$), position angle ($PA$), and the axis ratio ($q$). The fitting was done in two ways, similar to our previous study \citep{ahn17}. In short, we fitted while allowing all of the above parameters to vary, henceforth referred to as the ``free'' fit. The initial fits showed an isophotal twist between the individual S\'ersic profiles. However, the axis ratios of the outer profiles were nearly circular ($q$ $\sim$ 0.99). Furthermore, two of the three dynamical modeling techniques are restricted to axisymmetric potentials and thus, do not allow for isophotal twists. To enable comparison between all three techniques, we fixed the axis ratio of the outer profiles to be perfectly circular, and fitted the data again. Next, we fitted the data while fixing $R_e$, $n$, $PA$, and $q$ to the best-fit model from the other filter, which we call the ``fixed'' fit. For example, the fixed F814W fit contains all of the shape parameters from the best-fit F475W model, where only the total magnitude is varied. Since the only free parameter is the total magnitude, these fits provide a well defined color for each S\'ersic profile. The S\'ersic profiles used to create the default luminosity and mass models are shown in Figure~\ref{fig:sb}, and the parameters of the best fit models are shown in Table~\ref{tab:sersic}. We chose the default model to be the fixed F814W fit (outlined in bold in Table~\ref{tab:sersic}) due to 1.) the ability to accurately reproduce the surface brightness profile, 2.) it clearly provides the best fit to the color profile (discussed below), 3.) it provides a well defined color for each S\'ersic component. However, as discussed in Section~\ref{sec:JAM}, the choice of the luminosity and mass model produces a minor effect on the results of the dynamical models.

\begin{deluxetable}{clcc}[ht!]
  \tabletypesize{\scriptsize}
  \tablecaption{Best-fit S\'ersic Parameters}
  \tablewidth{0pt}
  \tablehead{\colhead{Component} & \colhead {Parameter}  & \colhead{F475W} & \colhead{F814W}}
  \startdata
  & $\chi^2$ & 1.045 & 1.356\\
  \hline
  & $m_{tot}$ & {\bf 17.27} & 16.24\\
  & $m_{tot}$ Fixed$^1$ & 17.58 & {\bf 16.01}\\
  & $R_e$ [$\asec$] & {\bf 0.21} & 0.26\\
  Inner & $R_e$ [$pc$] & {\bf 16.8} & 20.8\\
  & $n$ & {\bf 2.04} & 1.72\\
  & $q$ & {\bf 0.74} & 0.73\\
  & $PA$ [$^{\circ}$]$^{2}$ & {\bf -6.41} & -6.36\\
  \hline
  & $m_{tot}$ & {\bf 18.02} & 16.70\\
  & $m_{tot}$ Fixed$^1$ & 17.86 & {\bf 16.70}\\
  Middle & $R_e$ [$\asec$] & {\bf 0.50} & 0.35\\
  & $R_e$ [$pc$] & {\bf 40.0} & 28.0\\
  & $n$ & {\bf 0.80} & 5.23\\
  & $q$ & {\bf 1.00} & 1.00\\
  \hline
  & $m_{tot}$ & {\bf 18.84} & 16.95\\
  & $m_{tot}$ Fixed$^1$ & 18.20 & {\bf 17.78}\\
  Outer & $R_e$ [$\asec$] & {\bf 1.24} & 0.56\\
  & $R_e$ [$pc$] & {\bf 99.2} & 44.8\\
  & $n$ & {\bf 0.90} & 0.90\\
  & $q$ & {\bf 1.00} & 1.00\\
  \enddata
  \tablenotetext{1}{The ``Fixed'' magnitudes show the total magnitude when the shape parameters of the S\'ersic profiles are held fixed to the other filter.}
  \tablecomments{The PA orientation is N=0$^{\circ}$ and E=90$^{\circ}$. The default model (in bold) was constructed by fitting F475W first, allowing all parameters to be free. Then the shape parameters from the previous fit were held fixed and only $m_{tot}$ is fit for F814W. Next, we tested fitting F814W first, allowing all parameters to be free. Then the shape parameters from the previous fit were fixed and only $m_{tot}$ for F475W is fit. The numbers from the second approach are not bolded.}
  %\tablenotetext{2}{The PA orientation is N=0$^{\circ}$ and E=90$^{\circ}$.}
  %\tablenotetext{3}{The default model (in bold) was constructed by fitting F475W first, allowing all parameters to be free. Then the shape parameters from the previous fit were held fixed and only $m_{tot}$ is fit for F814W. Next, we tested fitting F814W first, allowing all parameters to be free. Then the shape parameters from the previous fit were fixed and only $m_{tot}$ for F475W is fit. The numbers from the second approach are not bolded.}
  \label{tab:sersic}
  \end{deluxetable}

The most massive (M $> 10^7$ $M_\odot$) UCDs have been found to consist of two components: a dense central component, and a more diffuse extended component, as shown by the two-component profile fits in previous studies \citep{evstigneeva07, evstigneeva08, chilingarianmamon08, strader13, ahn17,voggel18,afanasiev18}.  Shown in cyan in Figure~\ref{fig:sb}, a two component S\'ersic profile provides a suitable fit to the data within the central 2.5$\asec$. However, a much better fit out to 3$\asec$, was obtained by allowing a three-component S\'ersic profile, shown in Figure~\ref{fig:sb} as yellow.  The outer component is rounder, and as we will see below appears to be bluer than the inner components.  The inner S\'ersic components can also be replaced by a single King profile; a two-component King + S\'ersic profile provides an equally suitable fit as the three-component S\'ersic out to 3$\asec$.  The central King model component in this fit has a core radius $R_c = 0.05\asec$ (4~pc) and a concentration parameter $c = 1.42$.  We stick with our three-S\'ersic model fit for the remainder of the paper due to the ease of transforming this model into our mass models.

  The simulations of \citet{pfeffer13} show that when galaxies are stripped into UCDs, they show a two-component structure with the inner component consisting of the galaxies' nuclear star cluster (NSC) surrounded by the remains of the tidally stripped galaxy.   Therefore the multiple component model that is required to fit the surface brightness profile (regardless of the model choice) provides evidence that M59-UCD3 is likely a tidally stripped remnant.

The total luminosity in each band (within the central 3$\asec$) and effective radius calculated from the unconvolved S\'ersic profiles are found to be $L_{F814W} = 1.18-1.19 \times 10^8$ $L_{\odot}$, $L_{F475W} = 6.1-6.2\times 10^7$ $L_\odot$, and $R_e = 26-29$ pc or $0.32-0.36\asec$, respectively. Here the ranges are quoted based on the total luminosity and effective radius calculated across all free and fixed models. Our effective radii are consistent with previous measurements. However, our calculated luminosity is slightly lower than the previously estimated $L_g = 9.5 \times 10^7$ $L_\odot$ \citep{liu15,sandoval15}. We attribute these deviations in the luminosity to our deeper {\it HST} imaging data, where previous studies were limited to ground based data only. Each of the best-fit S\'ersic profiles were then parameterized by a multi-Gaussian expansion \citep[MGE:][]{emsellem94,cappellari02}, using the MGE\_FIT\_1D fitting method and code\footnote{\label{capnote}\url{http://purl.org/cappellari/software}} of \citet{cappellari02} for use in the dynamical modeling.

Creating dynamical models that reproduce the observed kinematics requires a luminosity and mass profile.  To determine the mass profile we make use of our dual filter {\it HST} data to test for the presence of stellar population variations. M59-UCD3 shows a complicated color profile that varies by 0.25 $mags$ within the central 3$\asec$, as shown in Figure~\ref{fig:color}. Here, the diamonds represent the data, where the error bars are calculated from our estimated systematic background effects, discussed above. Solid lines represent the convolved model and dashed lines represent the unconvolved model. The colored lines show whether the parameters of the S\'ersic fits were independent (black), or fixed (blue and red). Blue lines indicate that the shape parameters from the F475W filter were held fixed to the F814W filter, and red lines are vice versa. It is clear, from the unconvolved models, that the bluest colors near the center are due to PSF effects and there is a gradient toward bluer colors at larger radii. This is unique when compared to other UCDs, which generally show either no overall trend or a gradient toward redder colors are larger radii \citep{evstigneeva08, chilingarianmamon08, seth14, janz15,ahn17}.

The unconvolved fixed models (dashed blue and red lines in Figure~\ref{fig:color}) provide a well defined color for the three S\'ersic components. However, the fixed F814W S\'ersic shape parameters (red lines) provide a better fit to the color profile, which motivated our choice of the default model. Here, the inner component color is F475W $-$ F814W = 1.26, the middle component is F475W $-$ F814W = 1.32, and the outer component color is F475W $-$ F814W = 1.06. We used these colors combined with the \citet{bruzual03} Padova 1994 simple stellar population (SSP) models and corresponding code\footnote{\url{http://software.astrogrid.org/p/cea/latest/cec/config/galaxev/GALAXEV.html}}, assuming solar metallicity and a Chabrier IMF, to determine the mass-to-light ratio ($M/L$). The default \citet{bruzual03} color tables and corresponding $M/L$s do not include our {\it HST} filter set. Therefore, we downloaded the filter transmission curves and reran the composite stellar population model code. We found the inner, middle and outer $M/L_{F814W,*}$ to be $2.5 \pm 0.4$, $2.9 \pm 0.4$, and $1.0 \pm 0.1$, with corresponding ages of 9.9, 13.7, and 2.8 Gyrs, respectively. Here, the error bars are calculated assuming an uncertainty of $\pm0.05$ $mags$ in our color determinations. We also determined the total mass by multiplying the total luminosity with the corresponding $M/L$ for each S\'ersic component. We found the inner component total stellar mass to be $17.3 \pm 2.8 \times 10^7$ $M_\odot$, middle component to be $10.6 \pm 1.5 \times 10^7$ $M_\odot$, and the outer component to be $1.2 \pm 0.1 \times 10^7$ $M_\odot$. The mass density profile was then determined by multiplying the luminosity model MGEs by their corresponding $M/L$s. We note that to test the systematic effects of our choice of the default mass profile, we also determined two additional mass density profiles by 1.) computing the $M/L$ from the color of the free S\'ersic profile fits (black lines in Figure~\ref{fig:color}) at the FWHM of each Gaussian component in the MGE luminosity profile and 2.) by assuming a mass-follows-light model, which is equivalent to creating a mass density profile where all of the S\'ersic components have been scaled by the flux weighted $M/L$ described below. The latter test was motivated by the bluer outer most component, which could also be interpreted as a metal-poor old stellar population. As discussed in Section~\ref{sec:JAM}, this results in a $M/L_{F814W,*} = 1.9$, which is intermediate to our mass-follows-light model.

The luminosity is used to calculate the center of each kinematic bin and create the observed kinematic field. Since our kinematic data were taken in the $K$-band, we determined a luminosity MGE in that band using the color profiles and SSPs. This was accomplished by creating a color-color diagram of F814W $- K$ vs. F475W $-$ F814W from the SSP models. We then used our derived color to infer the F814W $- K$ color. For each S\'ersic component we found these colors to be F814W $- K$ =  2.28, 2.32, and 2.12 for the inner, middle, and outer components, respectively. These colors lead to a scale factor in the luminosity surface density for each component of 2.59 (inner), 2.69 (middle), and 2.24 (outer). These scale factors were multiplied by the luminosity profile for each component to make our $K$-band MGEs used in the dynamical models. The best-fit model MGE, for our default model, is shown in Table~\ref{tab:mge}.    

\begin{deluxetable}{cccc}[ht!]
  \tabletypesize{\scriptsize}
  \tablecaption{Default Multi-Gaussian Expansion (MGE) used in the dynamical modeling.}
  \tablewidth{0pt}
  \tablehead{\colhead{$Mass$ ($M_{\odot}pc^{-2}$)$^1$} & \colhead{$I_K$ ($L_\odot pc^{-2}$)$^2$} & \colhead{$\sigma (\asec)$} & \colhead{$q$}}
  \startdata
  294397. & 304328. & 0.001 & 0.74 \\%& -0.11\\
  361060. & 373239. & 0.004 & 0.74 \\%& -0.11\\
  370914. & 383426. & 0.011 & 0.74 \\%& -0.11\\
  308330. & 318732. & 0.027 & 0.74 \\%& -0.11\\
  199887. & 206630. & 0.059 & 0.74 \\%& -0.11\\
  99196.3 & 102542. & 0.115 & 0.74 \\%& -0.11\\
  36071.5 & 37288.3 & 0.209 & 0.74 \\%& -0.11\\
  9451.14 & 9769.96 & 0.356 & 0.74 \\%& -0.11\\
  1737.86 & 1796.49 & 0.576 & 0.74 \\%& -0.11\\
  210.794 & 217.905 & 0.903 & 0.74 \\%& -0.11\\
  11.8194 & 12.2182 & 1.459 & 0.74 \\%& -0.11\\
  \hline
  2547.19 & 2355.68 & 0.055 & 1.00 \\%& -56.78\\
  6801.10 & 6289.76 & 0.180 & 1.00 \\%& -56.78\\
  8380.02 & 7749.98 & 0.355 & 1.00 \\%& -56.78\\
  3882.25 & 3590.37 & 0.541 & 1.00 \\%& -56.78\\
  427.301 & 395.175 & 0.738 & 1.00 \\%& -56.78\\
  \hline
  44.2053 & 101.759 & 0.073 & 1.00 \\%& -70.25\\
  116.224 & 267.543 & 0.263 & 1.00 \\%& -70.25\\
  173.015 & 398.277 & 0.601 & 1.00 \\%& -70.25\\
  128.835 & 296.574 & 1.050 & 1.00 \\%& -70.25\\
  40.5159 & 93.2663 & 1.566 & 1.00 \\%& -70.25\\
  3.46218 & 7.96984 & 2.175 & 1.00 \\%& -70.25\\
  \enddata
  \tablenotetext{1}{The $M/L$ was used to determine the mass profiles. These methods are described in Section~\ref{hstphoto}.}
  \tablenotetext{2}{The luminosity MGEs were created in the $K$-band, which required an assumption on the absolute magnitude of the sun. We assumed these values to be 4.523 in F814W and 3.29 in $K$ taken from http://www.baryons.org/ezgal/filters.php. See Section~\ref{hstphoto} for a more detailed explanation of how the luminosity profile was derived.}
  \tablecomments{The PA adopted for all of the dynamical models was -6.41$^{\circ}$ where N=0$^{\circ}$ and E=90$^{\circ}$. The horizontal lines separate the individual S\'ersic models}
  %\tablenotetext{3}{The PA adopted for all of the dynamical models was -6.41$^{\circ}$ where N=0$^{\circ}$ and E=90$^{\circ}$}
  %\tablenotetext{4}{The horizontal lines separate the individual S\'ersic models}
  \label{tab:mge}
  \end{deluxetable}

Finally, the SSP models and color profiles were used to determine a flux weighted average $M/L$. We computed this by determining the flux within the central 3$\asec$ from model images of the inner, middle, and outer S\'ersic profiles and then weighting the individual $M/L$s calculated above by their corresponding flux. We found the average $M/L_{F814W,*}$ to be $2.47 \pm 0.25$ (assuming $\sim$10\% uncertainties similar to individual $M/L$s). We calculated the overall F475W $-$ F814W color to be 1.26. Using the SSP models, described above, we estimate $V -$ F814W = 0.86 \citep{bruzual03}. This corresponds to $M/L_{V,*}$ = $4.2\pm 0.4$, which is consistent with previous results \citep{liu15}.  We note that the \citet{mieske13} eq.~4 estimate of $M/L_{V,*}$ for [Fe/H]=-0.01 is 4.07 based on the mean between the \citet{bruzual03} and \citet{maraston05} model predictions. 

\subsection{Kinematic Derivation} \label{gemspec}

Our spectroscopic data were obtained with the Near-Infrared Integral Field Spectrometer (NIFS) mounted on the Gemini North telescope using {\it Altair} laser guide star adaptive optics \citep{mcgregor03, herriot00, boccas06}. The observations were taken on the nights of May 4th, 5th, and 6th, 2015, in the $K$ band at wavelengths from $2.0$ $\mu m$ to $2.4$ $\mu m$. Gemini/NIFS data are taken in $0.1 \asec \times 0.04\asec$ pixels over a 3$\asec$ field of view with a spectral resolution of $\frac{\lambda}{\delta \lambda} \sim 5700$ ($\sigma_{\rm inst} = 22$~km~s$^{-1}$).

Our data were reduced following the same procedure as our previous studies, using the Gemini version 1.13 IRAF package \citep{seth14, ahn17}. To summarize, first, arc lamp and Ronchi mask images were used to determine the spatial and spectral geometry of the images. Next, the spectra were sky subtracted, flat-fielded, had bad pixels removed, and split into long-slit slices. Finally, the spectra were corrected for telluric absorption with an A0V telluric star taken on the same night using the NFTELLURIC procedure. The final data cube consisted of eight (five on May 4th, one on May 5th, and two on May 6th) 900s on-source exposures taken in either an object-sky-object or object-sky sequence. The object exposures were dithered to give independent sky measurements for each exposure and to improve the signal-to-noise ratio ($S/N$). We created our own version of NIFCUBE and NSCOMBINE to rebin the spectra to a $0.05 \asec \times 0.05 \asec$ pixel scale, and enable error propagation of the variance spectrum. Finally, the spectra were combined using our IDL program which centroids the nucleus and rejects bad pixels based on the nearest neighboring pixels.

The kinematic PSF was determined by convolving a Gauss+\citet{moffat69} function with an {\it HST} model $K$ band image to match the surface brightness in the kinematic data cube. The {\it HST} model $K$ band image was derived following the same procedure outlined in Section~\ref{hstphoto} where we inferred the F814W $- K$ color using our derived color in each pixel combined with the SSP models. This model image was then convolved with a Gauss+Moffat function and fitted to the NIFS continuum image using the MPFIT2DFUN IDL code\footnote{\url{http://purl.com/net/mpfit}} \citep{markwardt09}. We note that in order to quantify the systematic effects of our PSF determination on the dynamical models, we also convoled the {\it HST} model image with a double Gaussian function. In this model PSF the central Gaussian contained 58\% of the light with a FWHM of 0.201$\asec$ and the outer Gaussian contained 42\% of the light with a FWHM of 0.894$\asec$. Furthermore, we also determined a PSF where we fixed the central Gaussian FWHM in our Gauss+Moffat model to be the diffraction limit ($FWHM = 0.07 \asec$) of the NIFS instrument. In this diffraction limited PSF the central Gaussian contained 24\% of the light, while the Moffat function contained the remaining 76\% with a FWHM of 1.08$\asec$. The effects of these PSFs are discussed in section~\ref{sec:JAM}. In our best-fit model function the central Gaussian was found to contain 35\% of the light with a FWHM of 0.165$\asec$. The Moffat function contained the remaining 65\% of the light with a FWHM of 1.08$\asec$. The one dimensional analytic Gaussian+Moffat PSFs were again fitted by Gaussians using the MGE\_FIT\_1D procedure of \citet{cappellari02}.

\begin{figure}[h!]
  \centering
  \includegraphics[trim={0cm 0cm 4cm 10cm},clip,scale=0.5]{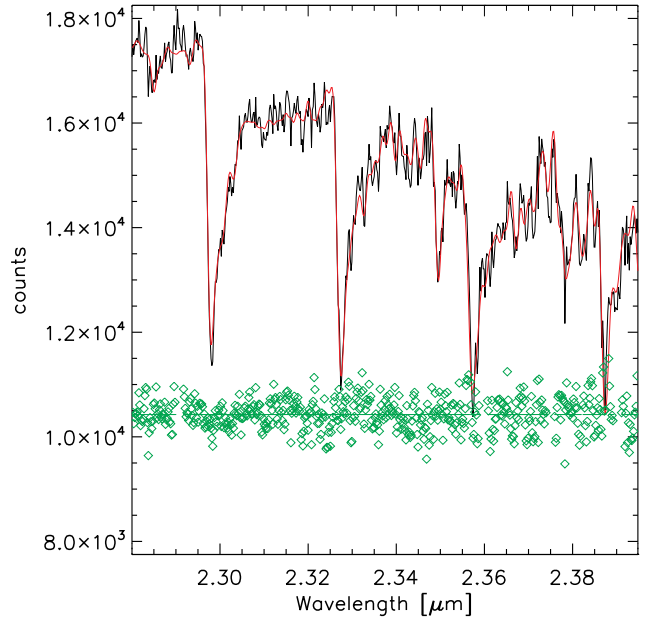}
  \caption{Integrated ($r < 0.75\asec$) spectrum of M59-UCD3 shown in black. The red line indicates the best kinematic fit, and residuals are shown in green. For visibility, the residuals are offset by $1.04\times10^4$ counts. The integrated dispersion was found to be $\sigma = 65.7 \pm 0.6$~km~s$^{-1}$ with a median $S/N = 57$ per pixel.}
  \label{fig:spectra}
\end{figure}

The kinematics were measured by fitting the CO bandhead region (2.28 $\mu m$ - 2.395 $\mu m$) to stellar templates using the IDL version of the penalized pixel algorithm pPXF, version 5.0.1\textsuperscript{\ref{capnote}} \citep{cappellari04,cappellari17}.
%XXX added reference
We convolved the high-resolution ($\frac{\lambda}{\Delta \lambda} = 45000$) \citet{wallace96} stellar templates with the line spread function (LSF), determined in each bin using sky exposures, before fitting. The bins were created using the Voronoi binning method and IDL code\textsuperscript{\ref{capnote}} to achieve $S/N =$ 25 per resolution element \citep{cappellari03}. However, beyond $0.5\asec$ the bins were remade using sectors to further increase the $S/N$ at the largest radii and enable identification of rotation signatures. These outer bins have $S/N \sim 20$, with the outermost four bins having $S/N \sim 12$. We fit the radial velocity (V), dispersion ($\sigma$), skewness ($h_3$), and kurtosis ($h_4$) to the data. The uncertainties associated with the determined kinematics were estimated using a Monte Carlo simulation. Gaussian random noise was added to each spectral pixel in each bin, and the kinematics were fitted again. The standard deviations of the fits were taken as the one sigma uncertainties. An example of the integrated ($r < 0.75\asec$) kinematic fits is shown in Figure~\ref{fig:spectra}. The barycentric systemic velocity and integrated ($r < 0.75\asec$) dispersion of M59-UCD3 were found to be $434.5 \pm 0.6$~km~s$^{-1}$ and $65.7 \pm 0.6$~km~s$^{-1}$, respectively. Previous measurements of the radial velocity have ranged from $373 \pm 18$~km~s$^{-1}$ to $447 \pm 3$~km~s$^{-1}$ \citep{sandoval15,liu15}. Our measured radial velocity is within the range of previous measurements, but outside the quoted uncertainties. Our velocity dispersion is significantly lower than previous measurements, which have ranged from $\sim70$~km~s$^{-1}$ (no quoted uncertainty) to $77.8 \pm 1.6$~km~s$^{-1}$ \citep{liu15,janz16}. We attempted to simulate the \citet{liu15} data by creating a Jeans Anisotropic Model (JAM) with the best-fit parameters (described below), a pixel size equivalent to their slit width, and seeing FWHM$=1.85\asec$. Using these values we found an integrated velocity dispersion of $67$~km~s$^{-1}$, which is still significantly lower than their measurement.

\begin{figure}[h!]
  \centering
  \includegraphics[trim={0 0cm 0 16cm},clip,scale=0.5]{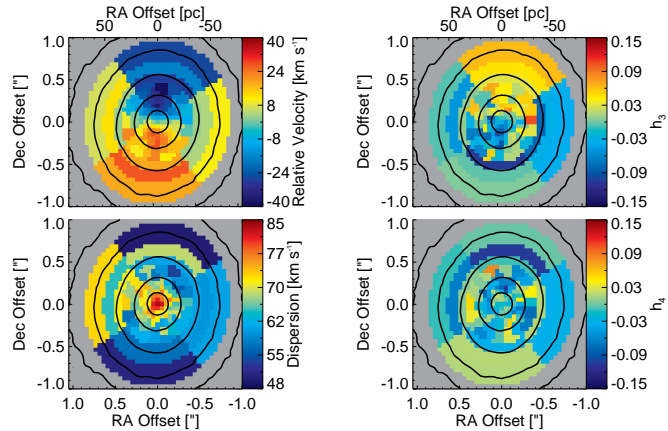}
  \caption{Full kinematic measurements of M59-UCD3, which include the radial velocity (top-left), dispersion (bottom-left), the skewness $h_3$ (top-right), and kurtosis $h_4$ (bottom-right). Black contours show the $K$ band continuum at 1 mag/arcsecond$^2$ intervals. The median 1$\sigma$ uncertainties are 3.9~km~s$^{-1}$ for the velocity, 3.7~km~s$^{-1}$ for dispersion, 0.04 for $h_3$, and 0.05 for $h_4$. Here, $h_3$ clearly shows the common anti-correlation with the velocity.}
  \label{fig:kin}
\end{figure}

  The kinematic maps for the four measured velocity moments are shown in Figure~\ref{fig:kin} {\bf and their corresponding values are shown in Table~\ref{tab:fullkin}}.  The velocity map shows clear rotation, and $h_3$ shows the common anti-correlation with the velocity. We also see a very definitive peak near the center of the dispersion map.  The dispersion map has some significant asymmetries, with higher values at larger radii to the East of the nucleus.

  \begin{deluxetable*}{cccccccccccccc}[ht!]
    \tabletypesize{\scriptsize}
    \tablecaption{Gemini/NIFS Kinematic Measurements}
    \tablewidth{0pt}
    \tablehead{\colhead{Bin} & \colhead{X rad} & \colhead{Y rad} & \colhead{Number} & \colhead{$S/N$} & \colhead{$\chi^2$} & \colhead{$v$} & \colhead{$v_{err}$} & \colhead{$\sigma$} & \colhead{$\sigma_{err}$} & \colhead{$h_3$} & \colhead{$h_{3,err}$} & \colhead{$h_4$} & \colhead{$h_{4,err}$} \\
      \colhead{Number} & \colhead{[pixels]} & \colhead{[pixels]} & \colhead{of Pixels} & \colhead{} & \colhead{} & \colhead{[km~s$^{-1}$]} & \colhead{[km~s$^{-1}$]} & \colhead{[km~s$^{-1}$]} & \colhead{[km~s$^{-1}$]} & \colhead{} & \colhead{} & \colhead{} & \colhead{}}
    \startdata
    ... & ... & ... & ... & ... & ... & ... & ... & ... & ... & ... & ... & ... & ... \\
    15 & -2.000 & -4.866 & 3 & 35.02 & 0.603 & 473.92 & 2.93 & 65.77 & 3.04 & -0.04 & 0.04 & -0.04 & 0.03 \\
    16 & -1.000 & -4.440 & 2 & 36.09 & 0.541 & 475.05 & 3.36 & 71.81 & 3.73 & -0.05 & 0.04 & -0.03 & 0.04 \\
    17 & 0.000 & -4.000 & 1 & 36.13 & 0.343 & 478.13 & 4.53 & 71.03 & 4.84 & -0.05 & 0.05 & -0.03 & 0.05 \\
    ... & ... & ... & ... & ... & ... & ... & ... & ... & ... & ... & ... & ... & ... \\
    
    \enddata
    \tablecomments{Only a portion of this table is shown here to demonstrate its form and content. A machine-readable version of the full table is available.}
    \label{tab:fullkin}
  \end{deluxetable*}
  
%  \begin{sidewaystable}[!htp]
%    \centering
%    \caption{Gemini/NIFS Kinematic Measurements}
%    \begin{tabular}{cccccccccccccc}
%      \hline \hline
%      Bin Number & X rad [$pixels$] & Y rad [$pixels$] & Number of Pixels & $S/N$ & $\chi^2$ & $v$ [km~s$^{-1}$] & $v_{err}$ [km~s$^{-1}$] & $\sigma$ [km~s$^{-1}$] & $\sigma_{err}$ [km~s$^{-1}$] & $h_3$ & $h_{3,err}$ & $h_4$ & $h_{4,err}$ \\
%      \hline
%      ... & ... & ... & ... & ... & ... & ... & ... & ... & ... & ... & ... & ... &... \\
%      15 & -2.000 & -4.866 & 3 & 35.02 & 0.603 & 473.92 & 2.93 & 65.77 & 3.04 & -0.04 & 0.04 & -0.04 & 0.03 \\
%      16 & -1.000 & -4.440 & 2 & 36.09 & 0.541 & 475.05 & 3.36 & 71.81 & 3.73 & -0.05 & 0.04 & -0.03 & 0.04 \\
%      17 & 0.000 & -4.000 & 1 & 36.13 & 0.343 & 478.13 & 4.53 & 71.03 & 4.84 & -0.05 & 0.05 & -0.03 & 0.05 \\
%      ... & ... & ... & ... & ... & ... & ... & ... & ... & ... & ... & ... & ... &... \\
%      \hline
%    \end{tabular}
    
%    \tablecomments{Only a portion of this table is shown here to demonstrate it%s form and content. A machine-readable version of the full table is available.}
%    \label{tab:fullkin}
%  \end{sidewaystable}
  
Using kinemetric fits\footnote{\url{http://davor.krajnovic.org/idl/}} \citep{krajnovic06}, we find the rotation amplitude peaks at $\sim$30~km~s$^{-1}$ at a radius of $0.2\asec$.  The $V/\sigma$ reaches $\sim$0.4 at this radius and stays constant at larger radii with both the dispersion and rotation decreasing outwards.  This is similar to the $V/\sigma$ of $0.3-0.5$ seen in other UCDs including M60-UCD1 and VUCD3 \citep{seth14,ahn17,voggel18}.  This level of rotation is higher than what is seen in any Milky Way globular clusters \citep[e.g.][]{kimmig15,kamann18}, and within the range of nearby galaxy nuclei with resolved observations \citep[e.g.][]{feldmeier14,nguyen17}.  

\section{Dynamical Modeling} \label{sec:model}
In this section we present our dynamical modeling techniques and their results. Here, we present three techniques: JAM, axisymmetric Schwarzschild modeling, and triaxial Schwarzschild modeling (but using an axisymmetric shape and mass model). Previous studies have shown that these techniques typically provide consistent results in estimating BH masses \citep[e.g.][]{verolme02,cappellari10,vandenbosch10,seth14,drehmer15,feldmeier-krause17,krajnovic18}. 

We note that our dynamical modeling effort and interpretation for M59-UCD3 is more complicated than for any previous results on BHs in UCDs \citep{seth14,ahn17,afanasiev18,voggel18}.  This is in part due to the high quality of the data, which enabled us to run triaxial Schwarzschild models in addition to the JAM models \citep[as we also did in M60-UCD1;][]{seth14}, and also due to the lower BH mass fraction in M59-UCD3 relative to previous UCD detections.  Our initial results showed the triaxial Schwarzschild and JAM codes gave inconsistent results for our BH mass, which led us to also run additional axisymmetric Schwarzschild models.  We were unable to fully resolve the differences here, but do our best to weigh the evidence from these models, and in Section~\ref{sec:summary} conclude that there is evidence for a $\sim$2\% mass fraction BH in M59-UCD3.  We discuss each modeling method individually below before reaching these conclusions.

%Section~\ref{sec:JAM} discusses the JAM models and the systematic effects of our choice of mass/luminosity and kinematic PSF models. Our preference to quantify the systematic effects with JAM was motivated by the computational resources needed for the Schwarzschild dynamical models. Section~\ref{sec:axischwmod} discusses the axisymmetric Schwarzschild models and Section~\ref{sec:triaxschwmod} explains our triaxial Schwarzschild models. In Section~\ref{sec:summary} we summarize the results and discuss their implications.

\subsection{Jeans Anisotropic Models (JAM)} \label{sec:JAM}
We use the JAM method to fit the $V_{RMS} = \sqrt{V^2 + \sigma^2}$ using the code\textsuperscript{\ref{capnote}} described in  \citet{cappellari08}. This technique solves the anisotropic Jeans equations under two general assumptions: (1) the velocity ellipsoid is aligned with the cylindrical coordinate system ($R,z,\phi$), (2) the anisotropy is constant and defined as $\beta_z = 1 - (\sigma_z/\sigma_R)^2$, where $\sigma_z$ is the velocity dispersion parallel to the rotation axis and $\sigma_R$ is the velocity dispersion in the radial direction \citep{cappellari08}. The primary input ingredients for the JAM models are the mass {\it and} luminosity density model (parameterized by MGEs as described in Section~\ref{hstphoto}).  These define the gravitational potential and the distribution of the tracer population which produces the observed kinematics. We use the JAM code to fit the following parameters: (1) the intrinsic axial ratio, which acts as a parametrization of the inclination angle ($q = \frac{\sqrt{q'^2-\cos^2{i}}}{\sin{i}} $, where $q$ is the parameter we sample over and $q'$ is the axis ratio of the flattest MGE), (2) the mass of a point-like black hole $M_{BH}$, (3) $\Gamma$, which parameterizes the $M/L$ relative to the best-fit stellar population estimate, and (4) the anisotropy parameter $\beta_z$ (see \S3.1 of \citealt{cappellari08} or \S4 of \citealt{ahn17} for a more detailed explanation of how these parameters are used).  For a given set of these four parameters, the JAM model generates observable model kinematic measurements that can be compared with the measured values.

For our JAM dynamical models we created a Markov Chain Monte Carlo (MCMC) simulation to fully sample parameter space. The parameters sampled include: the axial ratio, anisotropy parameter ($\beta_z$), $\Gamma$, and the mass of the BH. We used the {\it emcee} python package\footnote{\url{https://github.com/dfm/emcee}} developed by \citet{foreman-mackey13}, which is an implementation of the affine-invariant MCMC ensemble \citep{goodman10}. This algorithm uses a set of walkers to explore the parameter space. At each step, the result of the likelihood for each walker informs the next choice of model parameters to be evaluated. We ran all of our models with 200 steps per walker.

%Note that the JAM models enforce a simpler orbital structure compared to the Schwarzschild model formalism.  This tightens the constraints on e.g.~the BH mass relative to Schwarzschild models, and therefore we quote both the 1 and 3$\sigma$ uncertainties for all JAM model parameters as in our previous study \citep{ahn17}. %We also note that the JAM models only fit $V_{RMS}$, while the Schwarzschild models fit the full line-of-sight velocity distribution.

\subsubsection{Default Model}

\begin{figure}[h!]
  \centering
  \includegraphics[trim={0cm 0cm 0 0cm},clip,scale=0.3]{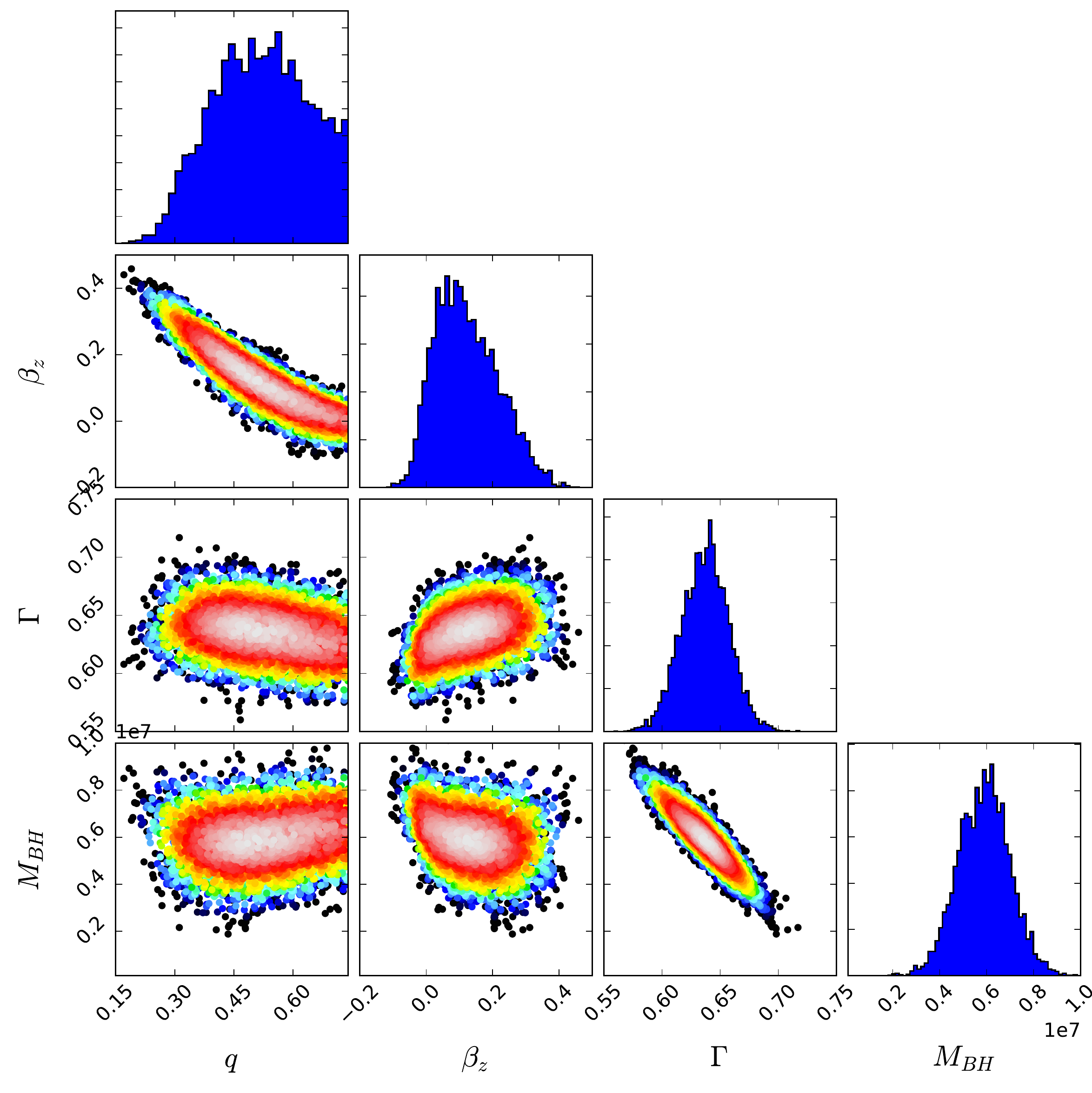}
  \caption{MCMC post burn-in phase distributions for our default model. The scatter plots show the projected two-dimensional distributions for each parameter. The histograms show the projected one-dimensional distribution. From top-left to bottom-right the panels show: the axis ratio, anisotropy parameter $\beta_z$, $\Gamma$, and $M_{BH}$.}
  \label{fig:mcmc}
\end{figure}

\begin{figure}[h!]
  \centering
  \includegraphics[trim={0.5cm 0cm 0 21cm},clip,scale=0.5]{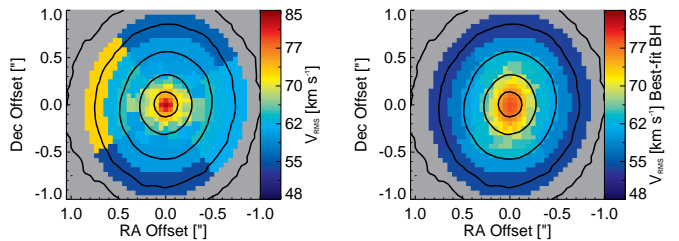}
  \caption{$V_{RMS}$ comparison between the smoothed data (left) and best-fit JAM model(right). Black contours show isophotes in the $K$ band stellar continuum.We note that the we smoothed the kinematic data in this figure for visual comparison only.}
  \label{fig:jamkin}
\end{figure}

For our default model, we used the $K$ band luminosity model to represent the tracer population, mass model determined from the fixed F814W surface brightness profile fit to represent mass distribution, and the best-fit Gauss+Moffat function PSF (all described in \S2). We ran our MCMC chain for this default model with 100 walkers for a total of 20000 steps. We consider the first 50 steps of each walker as the burn-in phase. Figure~\ref{fig:mcmc} shows the post burn-in phase distributions for this model. Here, the scatter plots show the two-dimensional distributions for each parameter with points colored according to their likelihood (white high and blue/black low). The histograms show the one-dimensional distributions for each parameter. We used the one-dimensional distributions to calculate the best-fit values and their corresponding uncertainties. The axis ratio is nearly unconstrained. The best-fit $\beta_z = 0.1 \pm 0.1$. We also see a slight degeneracy between $\beta_z$ and both $\Gamma$ and the mass of the BH ($M_{BH}$). The degeneracy seen here is less significant than we found in our previous study \citep{ahn17}, which we attribute to our higher quality kinematic data. The best-fit $\Gamma=0.64 \pm 0.02$ (which corresponds to $M/L_{dyn,F814W} = 1.58 \pm 0.05$ or $M/L_{dyn,V}=2.69 \pm 0.08$), and $M_{BH}=(5.9 \pm 1.1) \times 10^6$ $M_\odot$. The degeneracy between $M_{BH}$ and $\Gamma$ is expected, well known, evident across all of our dynamical modeling techniques (shown in Figure~\ref{fig:contour}, where blue contours represent the JAM models), and similar to what has been seen in previous studies \citep{seth14, ahn17,voggel18}. Here we quote the 1$\sigma$ uncertainties calculated from the 16th and 84th percentiles of the one-dimensional MCMC distributions. Due to the lack of orbital freedom in the JAM models we also quote the 3$\sigma$ uncertainties (0.2 and 99.8 percentiles), which encompass the systematic effects, discussed below. Quoting 3$\sigma$ errors we find $\beta_z = 0.1^{0.3}_{-0.2}$, $\Gamma = 0.64 \pm 0.06$ ($M/L_{dyn,F814W} = 1.58 \pm 0.15$, $M/L_{dyn,V} = 2.69 \pm 0.25$), and $M_{BH} = (5.9 \pm 3.1) \times 10^6$ $M_\odot$. Figure~\ref{fig:jamkin} shows the comparison of the smoothed data with the model $V_{RMS}$ calculated with the best-fit parameters. 

%  1 and 3$\sigma$ uncertainties, where the best-fit value is the 50th percentile, the 1$\sigma$ uncertainties are the 16th and 84th percentiles, and the 3$\sigma$ uncertainties are the 0.2 and 99.8 percentiles.} The axis ratio is nearly unconstrained. The best-fit $\beta_z = 0.1$ $^{+0.3}_{-0.2}$. We also see a slight degeneracy between $\beta_z$ and both $\Gamma$ and the mass of the BH ($M_{BH}$). The degeneracy seen here is less significant than what has been shown in previous studies, which we attribute to our high quality kinematic data \citep{ahn17}. The best-fit $\Gamma=0.64 \pm 0.06$ (which corresponds to $M/L_{dyn,F814W} = 1.58 \pm 0.15$ or $M/L_{dyn,V}=2.69 \pm 0.25$), and $M_{BH}=(5.9 \pm 3.1) \times 10^6$ $M_\odot$. The degeneracy between $M_{BH}$ and $\Gamma$ is expected, well known, and evident across all of our dynamical modeling techniques (shown in Figure~\ref{fig:contour}, where blue contours represent the JAM models), and similar to what has been seen in previous studies \citep{seth14, ahn17,voggel18}. Figure~\ref{fig:jamkin} shows the comparison of the data with the model $V_{RMS}$ calculated with the best-fit parameters. 

\subsubsection{Quantifying the Systematic Effects}
\begin{figure}[h!]
  \centering
  \includegraphics[trim={0 0cm 0 10cm},clip,scale=0.5]{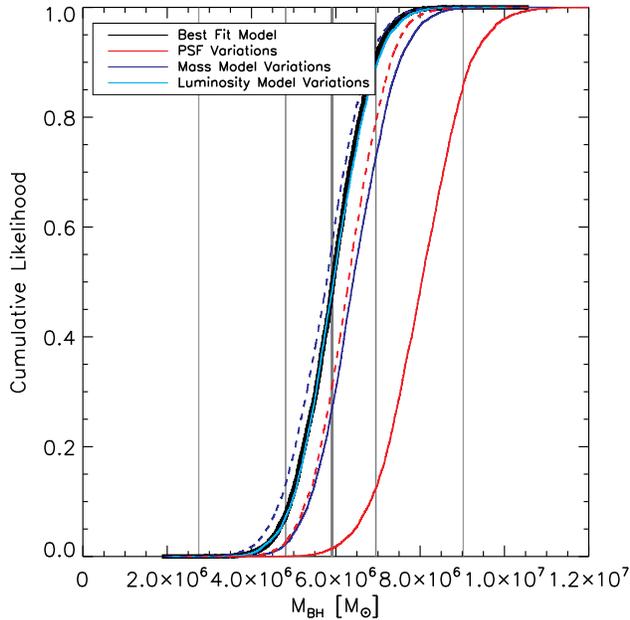}
  \caption{Cumulative likelihood of $M_{BH}$ from the JAM modeling. The black line represents the best-fit of the default model. The colored lines represent variations in the PSF (red), mass model (blue), and luminosity model (cyan). The grey vertical lines indicate the best-fit, 1, and 3$\sigma$ BH mass estimates based on the default model. See section~\ref{sec:JAM} for a detailed explanation the individual red, blue, and cyan lines.}
  \label{fig:cumulike}
\end{figure}

To quantify the systematic effects of our choice of the default model (discussed above) we also ran MCMC chains with 100 walkers and 20000 total steps. To demonstrate the effects of each systematic effect, we varied the PSF, mass model, and luminosity model one by one, while holding the other parameters fixed to our default model. The cumulative likelihood for all model variations is shown in Figure~\ref{fig:cumulike}. The black line represents the default model described above, with grey lines indicating the 1 and 3$\sigma$ uncertainties. The colored lines show variations in the PSF (red), mass model (blue), and luminosity model (cyan), which are explained in more detail below.

\begin{figure}
  \centering
  \includegraphics[trim={0.5cm 0cm 0 11cm},clip,scale=0.5]{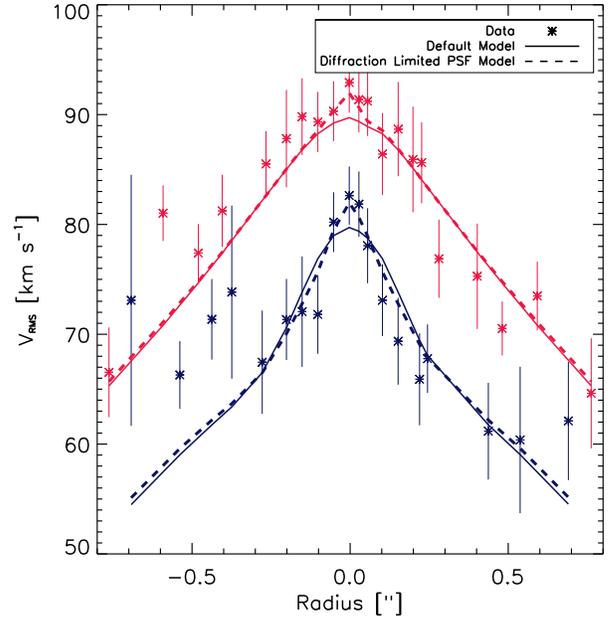}
  \caption{$V_{RMS}$ comparison between the data, best-fit default JAM model (solid with $M_{BH} = 5.9 \times 10^6$ $M_\odot$, $\Gamma = 0.64$), and best-fit diffraction limited PSF JAM model (dashed with $M_{BH} = 6.3 \times 10^6$ $M_\odot$, $\Gamma = 0.63$). Here, we take an elongated rectangular aperture one pixel wide along the semi-major axis (red) and the semi-minor axis (blue). The semi-major axis has been offset by 10~km~s$^{-1}$ for visibility. }
  \label{fig:difflim}
\end{figure}

The PSF is the largest source of systematic errors, seen in Figure~\ref{fig:cumulike} as the red lines.  Our default PSF was the best-fit model to the NIFS {\bf continuum} image from the $K$ band model. The solid line represents a double Gaussian model PSF, and the dashed line represents a Gauss+Moffat function model PSF where the central Gaussian FWHM is assumed to have a FWHM $= 0.07 \asec$, which is the diffraction limit of the NIFS instrument. In this model PSF, the Moffat function FWHM was left unchanged from the default PSF, but the corresponding weights of the Gauss and Moffat functions were recalculated to match the NIFS continuum. In all of the dynamical modeling techniques the model predictions are convolved with the kinematic PSF before comparison with the data (see Appendix~A of \citealt{cappellari08}). Therefore, our determination of the kinematic PSF is crucial in generating model observables that match the data. We tested the double Gaussian PSF as it is a common way to represent the adaptive optics PSF (but clearly did not fit the radial profile of the NIFS continuum as well in our case).  We also tested a PSF created by convolving of the F814W image as opposed to our $K$ band model and found very similar results (line not shown).  The diffraction limited PSF was tested due to our inability to match the central few pixel $V_{RMS}$ values. As shown in Figure~\ref{fig:difflim}, the best-fit parameters with the diffraction limited PSF allow the model $V_{RMS}$ values to better match the data near the center, but have minimal effect on the BH mass. Here we took an elongated rectangular aperture along the semi-major (red) and semi-minor (blue) axes. The solid lines represent the best-fit model $V_{RMS}$ with the default model, and the dashed lines represent the best-fit model with the diffraction limited PSF. Despite the better fit to the central data, we prefer our default model as the diffraction limited NIFS PSF, convolved with a model of the {\it HST} photometry, provides a worse fit to the surface brightness inferred from the NIFS datacube.

The blue lines in Figure~\ref{fig:cumulike} represent mass model variations with the default luminosity model and PSF. In this case, the solid line shows a mass model that was determined from the color of the free S\'ersic profile fits at the FWHM of each Gaussian in the MGE. This variation was motivated by the uncertainty in determining our mass profile from the fixed S\'ersic models. The dashed line shows our mass-follows-light model, which is equivalent to scaling all of the S\'ersic components by the flux weighted $M/L$ ($M/L_{F814W,*}=2.47$). This test was motivated by the bluer color of the outer component, which could also be due to an older, more metal-poor population at larger radii.  The de-extincted color of this component is redder than the \citet{bruzual03} models for metallicities below Z=0.004 ([Fe/H]$\sim-0.7$).  If we assume a model at that metallicity, we get a $M/L_{F814W,*}$=1.9;  this is closer (but still lower than) the inner component $M/L$s, and thus the resulting mass profile is intermediate between our constant $M/L$ and default model $M/L$s.  Figure~\ref{fig:cumulike} shows that these mass model variations provide $M_{BH}$ constraints within the 1$\sigma$ deviations from the default model and therefore our results do not depend critically on the stellar population variations.  

Finally, the cyan line represents a luminosity model variation where we used the default mass model and PSF, but the original fixed F814W luminosity MGE. The luminosity model variation makes the least difference. This is expected since the luminosity model is only used to determine the center of each kinematic bin and generate the observed kinematic field. Figure~\ref{fig:cumulike} shows that our choice for the default model is reasonable given that all of the model variations fall within the 3$\sigma$ error bars calculated from the default model likelihood.

\subsection{Axisymmetric Schwarzschild Models} \label{sec:axischwmod}
We fit the full line-of-sight velocity distribution (LOSVD) using an axisymmetric Schwarzschild orbit superposition model described in detail in \citet{cappellari06}. This three-integral dynamical modeling technique is based on Schwarzschild's numerical orbit superposition method \citep{schwarzschild79}, which has been shown to reproduce kinematic observations \citep{richstone88,rix97,vandermarel98}. This method assumes axisymmetry, which also requires the potential to not vary on the time scale required to sample the density distribution of an orbit. Since the orbital time scale within M59-UCD3 is $\sim 10^6$ yrs (assuming our effective radius and integrated dispersion), while the relaxation time is $\sim 10^{12}$ yrs and orbital time scale of M59-UCD3 around M59 is $\sim 10^8$ yrs, the potential is unlikely to vary during the orbital sampling period. The generality of this method has allowed it to become the standard for determining the mass of central BHs when high resolution kinematic data are available \citep[e.g.][]{verolme02,cappellari02a,gebhardt03,valluri05,shapiro06,vandenbosch06,nowak07,nowak08,krajnovic09,cappellari09}. However, the more general approach, which allows for triaxial systems, is described in \citet{vandenbosch08}, and discussed in Section~\ref{sec:triaxschwmod}.

%  Since the orbital time scale of M59-UCD3 around M59 is $\sim10^8$ yrs (assuming the projected radius quoted above, luminosity $L < R_{proj}$ from the S\'ersic profile fits of \citet{kormendy09}, dynamical M/L$\sim 6$) and the relaxation time of UCDs happens on much longer time scales, the potential is unlikely to vary during the orbital sampling period.} 

%Since the half-mass relaxation time scale of UCDs is $\sim10^{12}$ yrs (assuming the average stellar mass is $0.5 M_\odot$, and using our calculated half-light radius and integrated dispersion), the potential is unlikely to vary during the orbital sampling period. 

The full details of this method are described in \citet{cappellari06}. In short, this method consists of four steps. First, as with the JAM models, a stellar potential is created by deprojecting the mass model MGEs assuming an axisymmetric shape and stellar $M/L$. Second, a representative, dithered orbit library is constructed with even sampling across the observable sampling space (based on the three integrals of motion and the stellar potential). Next, the orbits are projected onto the observable space using sky positions, and taking into account the kinematic PSF and apertures (Voronoi bins, discussed in Section~\ref{gemspec}). Finally, the weights of each orbit are determined using a non-negative least squares fit \citep{lawson74}, which are co-added to reproduce the observed kinematics.

\begin{figure}[h!]
  \centering
  \includegraphics[trim={0.5cm 2cm 0 10cm},clip,scale=0.5]{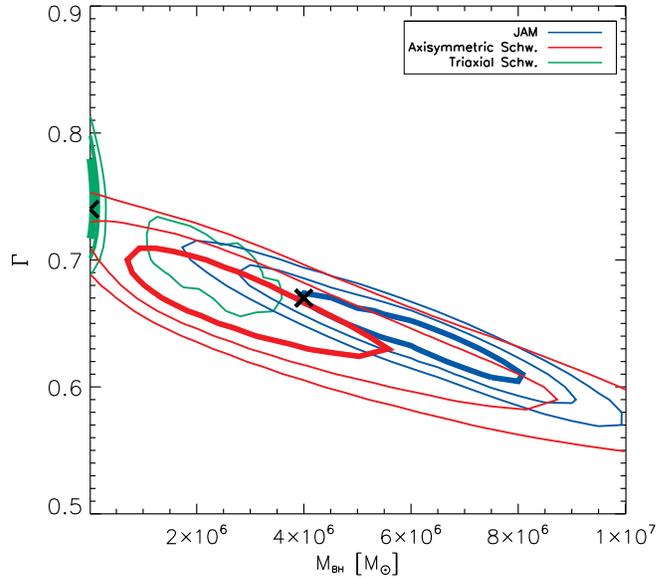}
  \caption{Contour plots showing all three modeling techniques. Here, the blue contours represent the JAM models, red are the axisymmetric Schwarzschild models and green are the triaxial Schwarzschild models. The black crosses denote the BH mass and $\Gamma$ values used to make the $V_{RMS}$ comparison plot shown in Figure~\ref{fig:compare1d}.}
  \label{fig:contour}
\end{figure}

For our models, we follow the approach outlined in \citet{krajnovic09}, with the only exception being that for M59-UCD3, we do not assume mass follows light. We used our default model, described above, to construct the mass and luminosity profiles. Here, the mass MGE is used to calculate the orbit libraries. For these models, we fixed the inclinations angle to be 85$^\circ$. This choice was arbitrary as the inclination angle has virtually no effect on the mass of the BH and $\Gamma$ (see Section~\ref{sec:JAM}). We created a grid of the two free parameters: $M_{BH}$ and $\Gamma$. The orbit libraries are constructed for each $M_{BH}$ at an expected $\Gamma$ and consist of 21x8x7x2 orbital bundles, which are composed of $6^3$ dithers (see \citealt{cappellari06}) This means there are 508032 total orbits, of which 2352 are free to vary to optimize the fit. It is not necessary to compute an orbit library for every ($M_{BH}$, $\Gamma$) combination because the orbit libraries can be scaled to match different $\Gamma$s. For our grid, we sampled 32 $M_{BH}$s between $2 \times 10^4$ $M_\odot$ and $1.2 \times 10^7$ $M_\odot$, and 47 $\Gamma$s between 0.43 and 0.89. The red contours in Figure~\ref{fig:contour} show the 1, 2 and 3$\sigma$ contour results for $\Gamma$ and $M_{BH}$. These contours were calculated from the LOESS smoothed $\chi^2$ distribution. Likewise, the best fits determined from the likelihood distribution are $M_{BH} = 2.5^{+1.8}_{-1.3} \times 10^6$ $M_\odot$, and $\Gamma= 0.67 \pm 0.03$ (1$\sigma$ uncertainties from the 16th and 84th percentiles).  We note that these models are consistent with no BH within the 3$\sigma$ contour.  

Figure~\ref{fig:contour} also shows the similar size and overlap between the 1$\sigma$ uncertainties from the axisymmetric Schwarzschild models (red contours) and the JAM models (blue contours). However, we note that these are formal errors and are smaller for JAM due to the reduced freedom of the model.

\subsection{Triaxial Schwarzschild Models} \label{sec:triaxschwmod}
Finally, we also fit the full LOSVD using the more general triaxial Schwarzschild models and corresponding code described in detail in \citet{vandenbosch08}. This model is also based on Schwarzschild's numerical orbit superposition method \citep{schwarzschild79}, but is not restricted to axisymmetry as described above. This method is implemented in a series of steps, similar to those described in Section~\ref{sec:axischwmod}. First, the stellar potential is created by deprojecting the mass model MGE, as described above. However, in the triaxial case, the viewing angles must be provided, which parameterize the intrinsic shape of the galaxy (see \textsection{3} of \citet{vandenbosch08}). Second, the initial conditions for each orbit library are found. These orbits must include all possible types of orbits that the potential can support \citep{thomas04,vandenbosch08}. Next, the orbits are integrated for a fixed period of time, while storing the projected properties on a grid for comparison with the data. These properties are convolved with the same PSF as the kinematic observations. Finally, the orbital weights are determined using a sparse quadratic programming solver from the GALAHAD library, which is capable of fitting the kinematics in a least-squares sense while also satisfying the mass constraints \citep{gould03}. 

For M59-UCD3, we provided an oblate axisymmetric shape by specifying the viewing angles ($\theta$,$\phi$,$\psi$) = (85, -49.99, 89.99) degrees. In the oblate limit, the model does not depend on $\phi$, while $\psi$ must be 90$^\circ$, and $\theta$ represents the inclination angle. Therefore, our model is nearly axisymmetric and seen at an inclination angle of $85^\circ$. The choice of the inclination angle was again arbitrary, but matches the axisymmetric Schwarzschild models. We sampled over all possible inclination angles and found consistent results. With this set up, the triaxial Schwarzcshild models are sampled in the axisymmetric limit, but still allow for all possible orbits in a triaxial potential. For the other two parameters, $\Gamma$, and $M_{BH}$, we created a grid similar to the axisymmetric Schwarzschild models described above. We sampled 47 $M_{BH}$s ranging from $5.5\times 10^3$-$2\times 10^7$ $M_\odot$, and 62 $\Gamma$s ranging from 0.43 to 1.04. The main results are shown as green contours in Figure~\ref{fig:contour}. Here, the best-fit $\Gamma = 0.75 \pm 0.06$ and we find $M_{BH}$ is consistent with no black hole.

%a 1$\sigma$ upper limit on $M_{BH}$ of $5\times10^4$~M$_\odot$.  

There is a clear disagreement on $M_{BH}$ between the JAM models/axisymmetric Schwarzschild models and these triaxial Schwarzschild models. To attempt to resolve these differences we explored a wide range of tests for our triaxial models including: fitting only the inner higher $S/N$ region, fitting sectors of the data, symmetrizing the kinematics, fitting only the radial velocity and velocity dispersion, adding various amounts of regularization, changing the total number of integrated orbits, and varying the input models and PSFs. In every test, the fitting results remained consistent. However, we note two interesting observations:\\(1) the green contours shown in Figure~\ref{fig:contour} show significant $\chi^2$ differences in the model in the region $M_{BH} < 2\times 10^5$ $M_\odot$.  At these masses, the BH sphere of influence is $<$0.002$\asec$, which is well below the diffraction limit of our NIFS data. This is clearly unphysical as the data cannot possibly constrain BH masses in this low-mass regime (i.e., the green contours are closed well below the diffraction limit of our instrument). We note that if we ignore models with $M_{BH} \lesssim 3 \times 10^5$ the triaxial model results become fully consistent with the JAM/axisymmetric Schwarzschild models.\\(2) We calculated the $\chi^2$ value for each of the model kinematic moments and $V_{RMS}$ independently. These values for two model BH masses are shown in Table~\ref{tab:chibyhand}, which shows the even kinematic moments and the $V_{RMS}$ favor a high-mass BH. However, the overall fit is clearly being driven by the odd velocity moments, especially the radial velocity. This is also unphysical, as the odd moments are supposed to provide virtually no constraints on the gravitational potential, as they have large freedom to vary, at fixed potential, to fit the data.  As discussed in Section~\ref{sec:summary}, comparing the $V_{RMS}$ profiles of the best-fit no black hole with a best-fit $M_{BH} \sim 4 \times 10^6$ $M_\odot$ shows a significantly better fit to the central pixels in the latter case. These observations lead us to speculate that the minimum $\chi^2$ at zero BH may be a numerical artifact, and to favor the results from the JAM and axisymmetric Schwarzschild models of a detectable SMBH. 

%Also, as discussed in Section~\ref{sec:summary}, comparing the dispersion profiles of the best-fit no black hole with a best-fit $M_{BH} \sim 4 \times 10^6 M_\odot$ shows a significantly better fit to the central dispersion in the latter case. These two observations lead us to favor the results from the JAM and axisymmetric models of a detectable SMBH.%believe a numerical problem is affecting the overall result and is therefore, not trustworthy.

\begin{deluxetable*}{cccccccc}[ht!]
  \tabletypesize{\scriptsize}
  \tablecaption{Calculations of Triaxial Schwarzschild Model Reduced $\chi^2$ Independently}
  \tablewidth{0pt}
  \tablehead{\colhead{$M_{BH}$ [$M_\odot$]} & \colhead{$\Gamma$} & \colhead{$\chi^2$ Total LOSVD} & \colhead{$\chi^2$ Vel Only} & \colhead{$\chi^2$ $\sigma$ Only} & \colhead{$\chi^2$ $h_3$ Only} & \colhead{$\chi^2$ $h_4$ Only} & \colhead{$\chi^2$ $V_{RMS}$}}
  \startdata
  $10^4$ & 0.75 & 0.765 & 1.005 & 0.875 & 0.791 & 0.498 & 0.799 \\
  $4 \times 10^6$ & 0.65 & 0.793 & 1.072 & 0.819 & 0.805 & 0.494 & 0.753 \\
  \enddata
  \label{tab:chibyhand}
  \end{deluxetable*}

\subsection{Summary of Dynamical Results} \label{sec:summary}

In summary, we detect a central massive black hole with the JAM dynamical models where the best-fit $M_{BH}$ and $\Gamma$ are $5.9 \pm 1.1 \times 10^6$ $M_\odot$ and $0.64 \pm 0.02$, respectively. With the axisymmetric Schwarzschild models we find the best-fit $M_{BH} = 2.5^{+1.8}_{-1.3} \times 10^6$ $M_\odot$ and $\Gamma = 0.67 \pm 0.03$ (1$\sigma$ uncertainties). Finally, with the triaxial Schwarzschild models we find the results are consistent with no black hole and $\Gamma =0.75$. However, the triaxial models show a small region that overlaps with the JAM/axisymmetric Schwarzschild models at the 3$\sigma$ level.  

\begin{figure}[h!]
  \centering
  \includegraphics[trim={0.5cm 0cm 0 11cm},clip,scale=0.5]{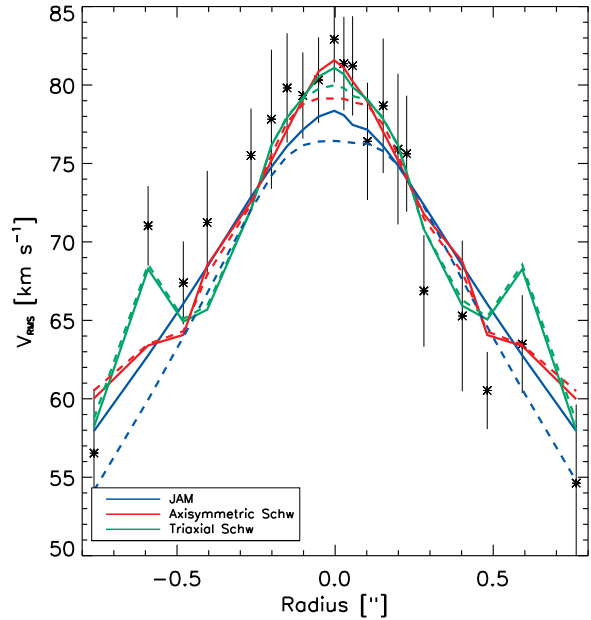}
  \caption{Black points show a rectangular aperture along the semi major axis. The solid lines represent a model $\sim 4 \times 10^6$ $M_\odot$ BH with $\Gamma = 0.67$ for the JAM (blue), axisymmetric Schwarzschild model (red), and triaxial Schwarzschild model (green). The dashed line represents a $\sim 10^4$ $M_\odot$ BH with $\Gamma = 0.74$ using the same colored convention described above. Note that these $M_{BH}$, $\Gamma$ combinations are not the best-fit model from any of the dynamical models. This choice is arbitrary, and is for visual comparison between a low and high mass BH only.}
  \label{fig:compare1d}
\end{figure}

Despite the variations in the dynamical modeling results, all of the models provide better fits to the $V_{RMS}$ data with a BH mass between $ 2-6 \times 10^6$ $M_\odot$.  This is particularly true in the central pixels, as shown in Figure~\ref{fig:compare1d}. Here, we show a $V_{RMS}$ model comparison for all of the dynamical modeling techniques along the semi-major axis. The colored lines show the JAM (blue), axisymmetric Schwarzschild models (red), and triaxial Schwarzschild models (green) best-fit parameters for two hypothetical $M_{BH}$, $\Gamma$ combinations, shown as X's in Figure~\ref{fig:contour}. In this case, we show a $M_{BH} \sim 4 \times 10^6$ $M_\odot$ with $\Gamma = 0.67$ as solid lines and $M_{BH} \sim 10^4$ $M_\odot$ with $\Gamma = 0.74$ as dashed lines. It is clear from this comparison plot that the high mass BH is favored in the $V_{RMS}$ profile for all of the dynamical modeling techniques, especially near the center where we expect the effects of a central massive BH are the most significant.

The results of the dynamical modeling techniques show that we cannot constrain the lower limit of the mass of a central massive BH. However, the better fits to the central $V_{RMS}$ profiles provides evidence in favor of a detectable BH mass.  Furthermore, the JAM and axisymmetric Schwarzschild models are nearly consistent at the 1$\sigma$ level. By combining the 1$\sigma$ confidence levels of the JAM and axisymmteric Schwarzschild models, we suggest the  BH mass in M59-UCD3 is $M_{BH} = 4.2^{+2.1}_{-1.7} \times 10^6$~M$_\odot$. This estimate is based on the average of the best-fit JAM and axisymmetric Schwarzschild models, where the uncertainties from each model were added in quadrature.  We do the same for the best $\Gamma$ value to find $\Gamma=0.65 \pm 0.04$, which corresponds to an average $M/L_{F814W,dyn} =1.61 \pm 0.10$ and $M/L_{V,dyn}=2.73 \pm 0.17$.

Finally, we note that this study is, to our knowledge, the first time that a direct comparison has been made between these three dynamical modeling codes.  As noted at the beginning of this section, in general, comparisons of JAM and Schwarzschild modeling have found consistent results.  One interesting recent study by \citet{leung18} has compared both Schwarzschild and JAM models against circular velocities derived from molecular gas for 54 galaxies with CALIFA integral-field stellar kinematics. The study found that JAM and Schwarzschild recover consistent mass profiles, without evidence for systematic biases (their Fig.~D1). However, it also found that JAM recovers more reliable circular velocities than the Schwarzschild models in the large-radii regime, where the gas velocities are more reliable (their Fig.~8). Although the study was not specific to SMBHs, it shows that the reduced generality of the JAM method, with respect to Schwarzschild's, can lead to a more robust mass-profiles recovery from real observations.  The lack of flexibility could be leading to a more robust result here too, especially if the kinematic data includes any outliers that are not well described by their error bars.  Finally, we note that despite the disagreement in the BH mass, the overall the agreement between the models is quite good; apart from the triaxial Schwarzschild model $\chi^2$ minimum at zero BH mass, the confidence regions of all three models overlap in both $\Gamma$ and $M_{BH}$.

\section{Radio and X-ray Observations of UCDs} \label{sec:xrayradio}
An alternative method for inferring the presence of a SMBH in UCDs is via accretion, which produces X-ray and radio emission. X-ray emission alone is only suggestive, as low-mass X-ray binaries are common in dense stellar systems and can mimic the X-ray emission from a low-luminosity AGN. However, radio emission from low-mass X-ray binaries is not detectable at the distance of the Virgo Cluster, and hence is a more secure indication of a SMBH.

Here we consider the radio and X-ray emission from three massive UCDs around the Virgo galaxies M59 and M60: M59cO, M59-UCD3, and M60-UCD1, which all have dynamical evidence for SMBHs.  We note that no deep radio data exist for the other UCDs with evidence of SMBHs.

\subsection{Radio}

We obtained deep radio continuum data for M59 and M60 with the Karl G.~Jansky Very Large Array (VLA) as part of program 15A-091 (PI: Strader) in February and March 2015. All data were taken in B configuration and with C band receivers in 3-bit mode, split into subbands centered at 5 and 7 GHz, each with 2 GHz of bandwidth. Four 1.75 hr long blocks were observed, and in each block observations alternated between the two targets, giving 3.5 hr of observations (2.6 hr on source) per galaxy. The data were flagged and calibrated in AIPS using standard methods, then imaged with Briggs robust weighting \citep{briggs95}. The subband data were imaged separately (at central frequencies of 4.6 and 7.1 GHz after flagging) and together, at a mean frequency of 5.8 GHz.
The beam in the combined images is $1.33\arcsec \times 1.14\arcsec$.

\begin{figure}[h!]
  \centering
  \includegraphics[trim={1.5cm 0cm 1.5cm 1cm},clip,scale=0.25]{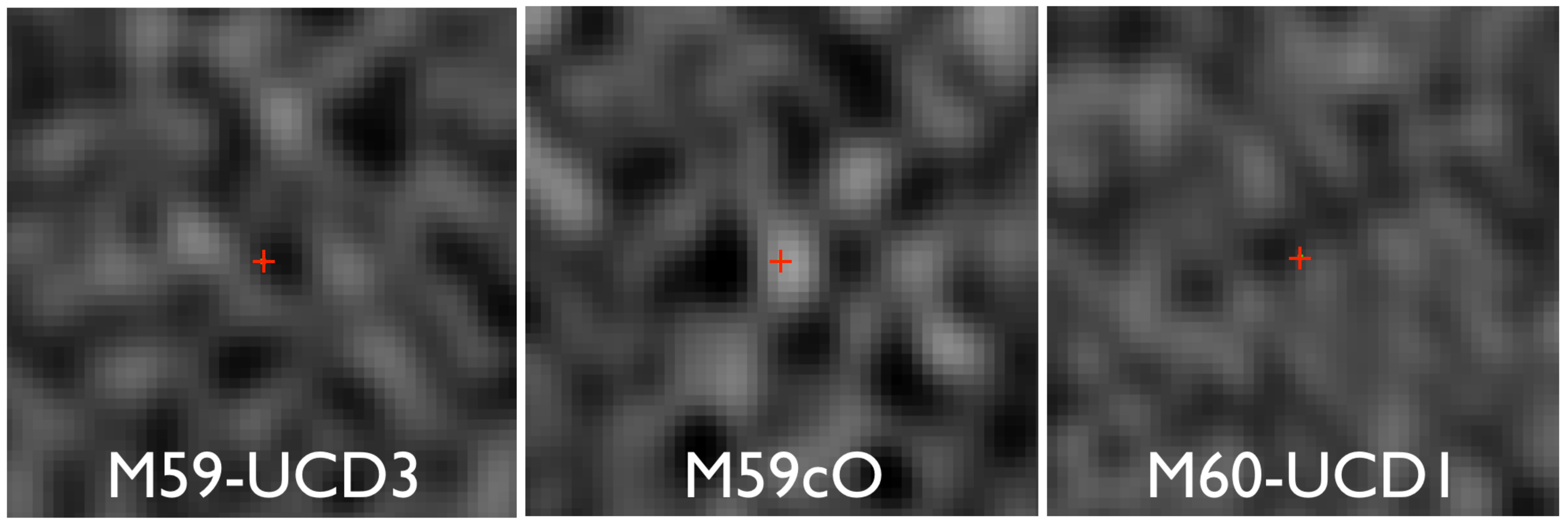}
  \caption{VLA images of a 10$\asec$ (800 pc) box around three massive UCDs with dynamical evidence for SMBHs: M59-UCD3, M59cO, and M60-UCD1. The red crosses mark the optical positions of the UCDs. M59cO has evidence for an associated radio source as discussed in the text, while M59-UCD3 and M60-UCD1 are not detected in the VLA images.}
  \label{fig:mosaic}
\end{figure}

M59-UCD3 is not detected in the individual subbands or in the combined image. The local rms noise in the region of M59-UCD3 is 2.6 $\mu$Jy bm$^{-1}$. Hence, we set a 3$\sigma$ upper limit of  $< 7.8 \mu$Jy bm$^{-1}$ (L $< 1.27 \times 10^{34}$) at a mean frequency of 5.8 GHz. M60-UCD1 is also undetected, with a local rms of 2.4 $\mu$Jy bm$^{-1}$ and a corresponding upper limit of  $< 7.2 \mu$Jy bm$^{-1}$ (L $< 1.17 \times 10^{34}$) at 5.8 GHz. In contrast, we do detect M59cO in the 4.6 GHz subband at a flux density of $10.8\pm3.8$ $\mu$Jy bm$^{-1}$ (L $= 1.75 \times 10^{34}$).  It is not detected in the 7.1 GHz image. The UCD is detected in \emph{Gaia} with a J2000 position of (R.A., Dec) = (12:41:55.334, +11:40:03.79), only $0.1\arcsec$ from the VLA position of the radio source in the 4.6 GHz image (R.A., Dec) = (12:41:55.331, +11:40:03.69). The astrometric match suggests that the radio emission, while faint, is indeed real and associated with M59cO. Here, the luminosity is the flux density in $\mu$Jy $\times 10^{-29} \times 4\pi R^2 \times 5\times 10^9$. These are all given at 5 GHz (i.e., assuming a flux density slope of $\alpha = 0$ (flat)). VLA mosaic images for these three UCDs is shown in Figure~\ref{fig:mosaic}.

\subsection{X-ray}

These UCDs have been studied in the X-rays using \emph{Chandra} by several previous authors \citep{luo13,strader13,pandya16,hou16},  but we revisit this analysis to ensure consistency. All our results are consistent with these past studies.  As noted in these previous studies, the X-ray emission from UCDs can be explained by low-mass X-ray binaries (LMXBs).   In fact the number of X-ray sources falls short of expectations based on globular cluster X-ray sources, but SMBH emission cannot be excluded \citep{pandya16,hou16}.  Given that 10$^{6-7}$~M$_\odot$ SMBHs do seem to be present in UCDs, if these are accreting at the typical Eddington ratios seen for early-type  galaxies \citep[$L_{bol}/L_{edd} \sim 10^{-6}$][]{ho09},
%XXXho09: http://adsabs.harvard.edu/abs/2009ApJ...699..626H
we would expect the UCDs to have detectable X-ray sources of $\sim$10$^{38}$ ergs/s.  As discussed further in the next section, the radio emission from LMXBs is much lower than that expected for emission from SMBHs and thus a detection of both X-ray and radio emission from a source would provide strong evidence for SMBH accretion.  

There are two separate observations of M59 (encompassing both M59-UCD3 and M59cO) and six observations of M60 that cover M60-UCD1; these are summarized in Table~4. We downloaded these observations from Chandra data archive and reprocessed them using  \texttt{CIAO} 4.9 and \texttt{CalDB} 4.7.6. We used a $1.5\arcsec$ extraction radius around each source, and measured the background in a larger nearby source-free area before normalizing the counts to the source extraction region size. We initially determined all counts in the 0.3--10 keV range for maximum sensitivity, but report results in the 0.5--10 keV range for appropriate comparison to the fundamental plane. For both galaxies we fix $N_H = 2\times10^{20}$ cm$^{-2}$ (taking extinction from \citet{schlafly11} and conversion from \citet{bahramian15}). All spectral extractions were performed with CIAO task specextract and spectral analysis was done using Xspec 12.9.1n \citep{arnaud96}. We assumed \citet{wilms00} abundances and \citet{verner96} absorption cross-sections. 

M59cO is not detected in the 2001 or 2008 observations. Assuming a power-law with $\Gamma = 1.5$, in the 2001 data we find a 95\% upper 0.5--10 keV unabsorbed flux limit of $< 5.3\times10^{-16}$ erg s$^{-1}$ cm$^{-2}$, equivalent to $L_X < 1.7 \times 10^{37}$ erg s$^{-1}$. The shorter 2008 data are less constraining and give a limit of $L_X < 1.4 \times 10^{38}$ erg s$^{-1}$ using the same assumptions. 

M59-UCD3 is detected at $> 2\sigma$ in the 2001 \emph{Chandra} data, with a 0.5--10 keV unabsorbed flux of $3.1^{+2.7}_{-1.7}  \times10^{-15}$ erg s$^{-1}$ cm$^{-2}$, equivalent to $1.0^{+0.9}_{-0.6}  \times10^{38}$ erg s$^{-1}$ (uncertainties are at the 95\% level). Unsurprisingly it is not detected in the factor of $\sim 5$ shorter 2008 data. In addition, it is located near a chip gap in the 2008 observations, which makes it difficult to determine a valid upper flux limit. Here, we have assumed Gehrels statistics for all of the upper limits \citep{gehrels86}.

M60-UCD1 is detected in all six observations. The total merged dataset, representing 308 ksec of \emph{Chandra} data, is deep enough to allow spectral fitting. After binning to 20 counts per bin, we fit the spectrum to a power-law in XSPEC using cstat, a modified version of the Cash statistic \citep{cash79}.\footnote{\url{https://heasarc.gsfc.nasa.gov/xanadu/xspec/manual/XSappendixStatistics.html}}. The best-fitting power law index is $\Gamma = 1.8^{+0.2}_{-0.3}$, consistent with the $\Gamma = 1.5$ value assumed. Hence, for consistency, we assume $\Gamma = 1.5$ for all the flux measurements for M60-UCD1.

The individual unabsorbed 0.5--10 keV fluxes for M60-UCD1 range over 1.8--$7.5 \times10^{-15}$ erg s$^{-1}$ cm$^{-2}$ ($L_X = 0.6$--2.4$\times 10^{38}$ erg s$^{-1}$, depending on the epoch. The average flux is $(3.3^{+0.8}_{-0.7} \times10^{-15}$ erg s$^{-1}$ cm$^{-2}$ ($L_X = 1.1^{+ 0.3}_{-0.2} \times 10^{38}$ erg s$^{-1}$). The individual and merged fluxes are listed in Table~\ref{tab:swift}. 
 
There is compelling evidence for X-ray variability of  M60-UCD1, but only at a single epoch: five of the six epochs are consistent with the mean flux, while one (Obs ID 12976) is $\sim 6\sigma$ higher compared to the mean flux. Due to the shorter exposure times and smaller number of epochs for M59-UCD3 and M59cO, we have no useful constraints on X-ray variability for these other sources.

\subsection{Fundamental Plane of BH Accretion}

We can combine X-ray and radio detections and non-detections described above with the dynamical black hole mass estimates to see if these observations are consistent with the fundamental plane of BH accretion \citep{merloni03,falcke04}.  We note that significant variability is seen in M60-UCD1, and long-term variability of low-luminosity AGN seems to be common \citep[e.g.][]{maoz05,hernandez-garcia14}, and in general UCD X-ray sources appear to be variable \citep{pandya16,hou16}.
%XXX hernandez-garcia14: http://adsabs.harvard.edu/abs/2014A%26A...569A..26H
%XXX maoz05: http://adsabs.harvard.edu/abs/2005ApJ...625..699M
Because our radio and X-ray observations are not contemperaneous, this variability adds to any intrinsic scatter present.

\begin{figure}[h!]
  \centering
  \includegraphics[trim={2cm 13cm 3cm 1cm},clip,scale=0.6]{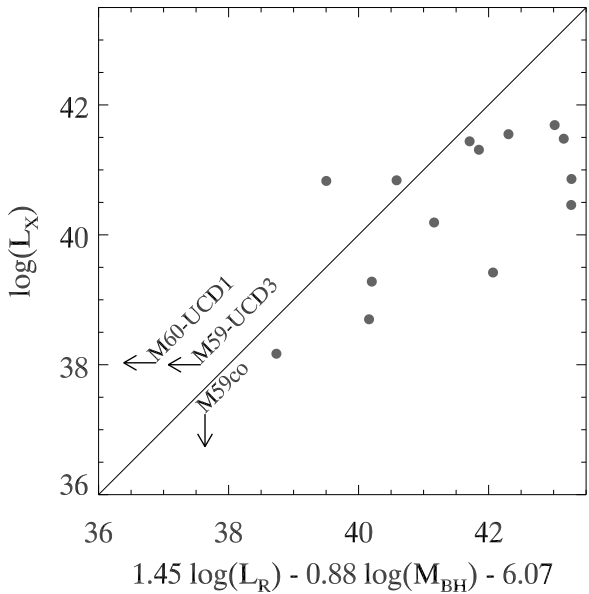}
  \caption{The fundamental plane for UCDs with known BHs from \citep{plotkin12}.  The radio upper limits in M60-UCD1 and M59-UCD3 and the X-ray upper limits in M59cO (arrows) are all below the predictions from the fundamental plane, but these differences are small relative to the scatter seen for similar quiescent black holes with dynamical BH mass determinations shown in gray from \citet{gultekin09b}.}
  \label{fig:fp}
\end{figure}

We use the fundamental plane of \citet{plotkin12}:\\
\begin{equation}
  \log(L_X)=(1.45 \pm 0.04) \log(L_R) - (0.88 \pm 0.06) \log(M_{BH}) - 6.07\pm1.10
\end{equation}
and plot the combinations of detections and upper limits for the three UCDs in Fig.~\ref{fig:fp}.  We find that the radio upper limits in M59-UCD3 and especially M60-UCD1 fall well below the radio luminosities expected for objects lying on the fundamental plane.  Similarly, the X-ray non-detection of M59cO is below the expectation based on its radio luminosity.  However, these non-detections do not provide strong constraints on whether an accreting BH is present.  This is because of the order of magnitude scatter in the radio luminosities relative to the fundamental plane observed in similar low-Eddington systems with known BH masses \citep{gultekin09b}, as well as the lack of simultaneous radio and X-ray observations.  Future, simultaneous detections of X-ray and radio emission in UCD BHs could provide important confirming evidence for SMBHs in these systems.
    %gultekin09b http://adsabs.harvard.edu/abs/2009ApJ...706..404G

Finally, we note that from our NIFS observations, we can also constrain the presence of hot dust emission in these systems.  This hot dust is typically thought to be an AGN accretion signature and is found to be quite luminous in both quiescent and actively accreting BHs \citep{seth10,seth10b,burtscher15}.  A clear correlation is seen between this emission and the X-ray and mid-infrared emission, although this correlation seems to depend significantly on AGN type \citep{burtscher15}.  Whether this correlation continues to lower luminosity AGN like those observed here is not yet clear.  %Using direct fitting of stellar templates plus a hot dust component, Dumont et al., {\em in prep} finds clear detections of compact hot dust emission at the centers of both M59-UCD3 and M60-UCD1.  The $K$ band luminosities are XXX and XXX respectively.  This data may provide additional evidence for the presence of SMBH accretion in these two X-ray detected systems.  

\begin{deluxetable*}{llccccc}[ht!]
  \tablecaption{Chandra X-ray Constraints and Measurements}
  \centering
  \tabletypesize{\scriptsize}
  %\tablecolumns{10} 
  %\tablewidth{0pt} 
  %\rotate
  \tablehead{
    \colhead{UCD ID} &
    \colhead{Obs. ID} &
    \colhead{Epoch Date}&
    \colhead{Epoch Date}&
    \colhead{Effective time} &
    \colhead{Flux} & 
    \colhead{Luminosity} \\
    \colhead{}&
    \colhead{}&
    \colhead{}&
    \colhead{(MJD)} &
    \colhead{(ksec)}&
    \colhead{($\times 10^{-15}$ erg s$^{-1}$ cm$^{-2}$)}&
    \colhead{($\times 10^{37}$ erg s$^{-1}$)} 
  }
  \startdata
  M59-UCD3 & 2068 &  &  & 24.8 & $3.1_{-1.7}^{+2.7}$ & $10.1_{-5.5}^{+8.8}$ \\
  M59-UCD3 & 8074 &  &  & 5.3 & \nodata & \nodata \\
  \hline
  M59cO & 2068 &  &  & 24.8 & $< 0.5$ & $< 1.7$ \\
  M59cO & 8074 &  &  & 5.3 & $< 4.4$ & $<14.3$ \\
  \hline
  M60-UCD1 & 785 & 2000-04-20 & 51654.148669 & 38.1 & $1.8_{-0.9}^{+1.3}$ & $5.9_{-2.9}^{+4.2}$ \\
  M60-UCD1 & 8182 & 2007-01-30 & 54130.521317 & 52.4 & $3.7_{-1.3}^{+1.7}$ & $12.1_{-4.2}^{+5.5}$ \\
  M60-UCD1 & 8507 & 2007-02-01 & 54132.122880 & 17.5 & $3.6_{-2.2}^{+3.7}$ & $11.7_{-7.2}^{+12.1}$ \\
  M60-UCD1 & 12976 & 2011-02-24 & 55616.730009 & 101.0 & $7.5_{-1.4}^{+1.6}$ & $24.4_{-4.6}^{+5.2}$ \\
  M60-UCD1 & 12975 & 2011-08-08 & 55781.313337 & 84.9 & $2.8_{-1.0}^{+1.2}$ & $9.1_{-3.3}^{+3.9}$ \\
  M60-UCD1 & 14328 & 2011-08-12 & 55785.067169 & 14.0 & $2.4_{-1.9}^{+3.3}$ & $7.8_{-6.2}^{+10.8}$ \\
  M60-UCD1 & all & \nodata       & \nodata      & 307.9 & $3.3_{-0.7}^{+0.8}$  & $10.8_{-2.3}^{+2.6}$ \\
  \enddata
  \tablecomments{All limits and uncertainties are at the 95\% level and over the energy range 0.5--10 keV. Luminosities assume a distance of 16.5 Mpc.}
  \label{tab:swift}

\end{deluxetable*}

\section{Discussion \& Conclusions} \label{sec:discussconclude}
In this paper we have presented the results of three separate dynamical models on the most massive UCD, M59-UCD3, and discussed the radio and X-ray observations of UCDs as a way to infer the presence of SMBHs. Detections of SMBHs in UCDs provide evidence that they are the tidally stripped remnants of once more massive galaxies \citep{seth14,ahn17,afanasiev18}. Furthermore, the effects of the central massive BH can explain the elevated dynamical-to-stellar mass ratios detected in the most massive UCDs \citep{mieske13}. Non-detection of BHs can suggest either that the object is not a stripped nucleus (and instead is a massive star cluster), or that the nucleus stripped lacked detectably massive central BHs  \citep{voggel18} .

\subsection{Summary of Main Results}
For our analysis, we combined adaptive optics assisted Gemini/NIFS kinematic data with high resolution {\it HST} imaging. The Gemini/NIFS data were used to determine the full LOSVD, which includes the radial velocity, velocity dispersion, skewness, and kurtosis. We found the integrated ($r < 0.75\asec$) barycentric radial velocity to be $V = 434.5 \pm 0.6$~km~s$^{-1}$ and velocity dispersion $\sigma = 65.7 \pm 0.6$~km~s$^{-1}$. The {\it HST} images were used to construct a mass density, and luminosity profile by fitting a PSF convolved triple component S\'ersic profile to the data. These models were used to calculate the total luminosity (within the central 3$\asec$), which we found to be $L_{F814W} = 1.19\times10^8$ $L_\odot$ and an effective radius of $0.34\asec$ (27~pc). The model fits suggest the outer component of the UCD is somewhat bluer than the central components, and we modeled this stellar population variation using SSP models. 
%The dual filter {\it HST} data were also used to test for the presence of multiple stellar populations. We accounted for these effects by creating a mass density profile from the luminosity density profile by multiplying the corresponding $M/L$s from the SSP models.

We combined these mass and luminosity density profiles  with the kinematic measurements to test for the presence of a central massive BH using three dynamical modeling techniques.  We summarize the results of this modeling in (Section~\ref{sec:summary}); our final conclusion is that M59-UCD3 appears to host a BH with a mass of $M_{BH} = 4.2^{+2.1}_{-1.7} \times 10^6$~M$_\odot$.
%while the best-fit $\Gamma = XXX$, corresponding to $M/L=XXX$ and a total mass for M59-UCD3 of XXX.  
We derive a best-fit $\Gamma$ from the JAM and axisymmetric Schwarzschild models of $0.65 \pm 0.04$, which corresponds to an $M/L_{F814W,dyn} = 1.61 \pm 0.10$ and $M/L_{V,dyn} = 2.73 \pm 0.17$.  Therefore, the total dynamical mass is $M_{dyn} = 1.9 \pm 0.1 \times10^8$ $M_\odot$.

%First, we fitted the $V_{RMS}$ using the JAM dynamical modeling technique. We found a black hole with 3$\sigma$ uncertainties to be $M_{BH} = 5.9 \times 10^6 M_\odot$, with a corresponding dynamical $M/L_{F814W} = 1.58 \pm 0.15$. Next, we fitted the full LOSVD with axisymmetric Schwarzschild models and found the best fit BH mass and dynamical $M/L$ to be $M_{BH} = 2.5^{+7.2}_{-2.4} \times 10^6 M_\odot$ and $1.65 \pm 0.25$, respectively. Finally, we fitted the full LOSVD with triaxial Schwarzschild models and found the results to be consistent with no BH. However, multiple tests point to a numerical issue in the modeling code, for which the solution is beyond the scope of this paper. Therefore, we prefer the solutions from the JAM and axisymmetric Schwarzschild models.

\subsection{Implications of a Central Massive BH}
\begin{figure}[h!]
  \centering
  \includegraphics[trim={0.5cm 0cm 0 10.5cm},clip,scale=0.5]{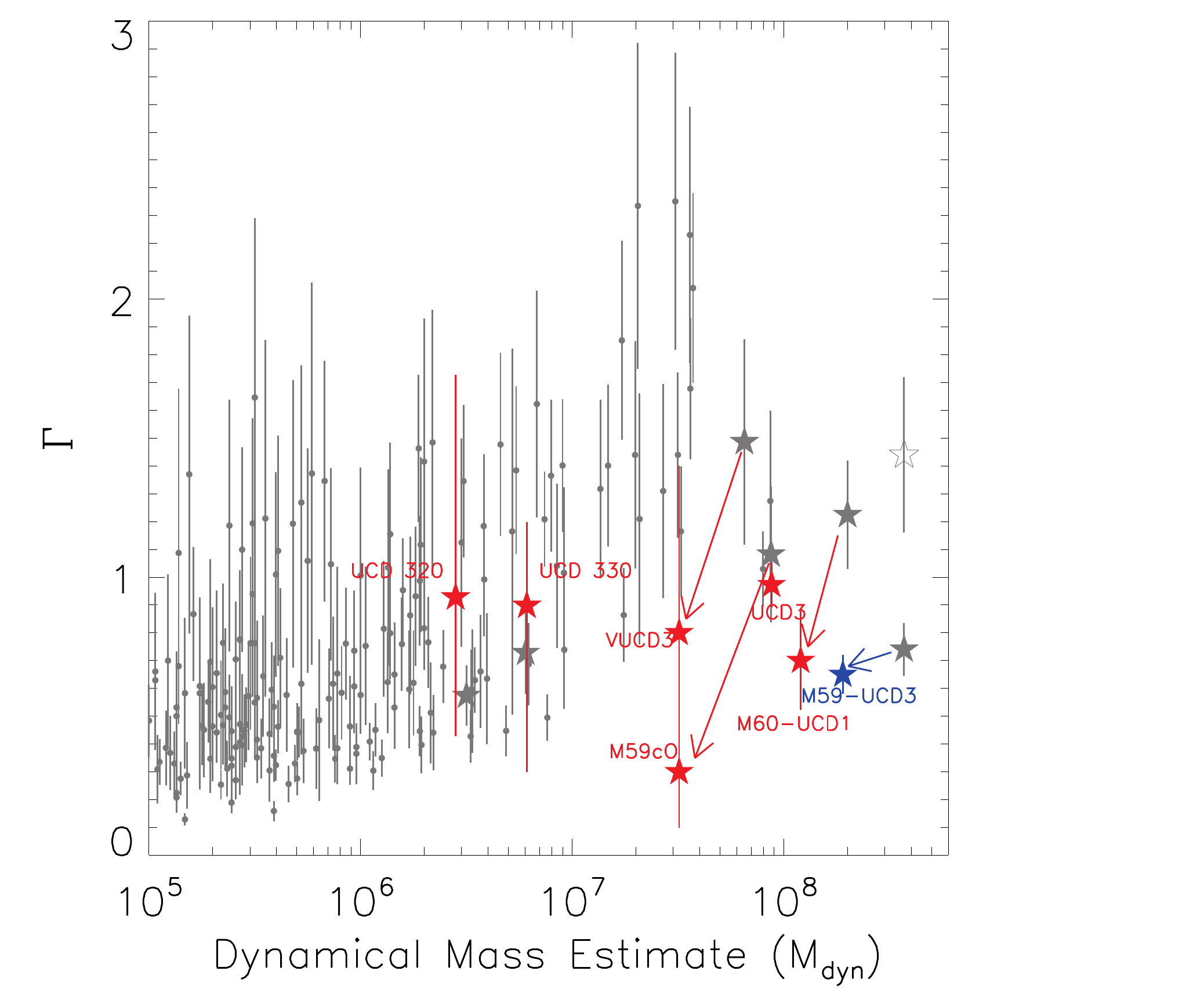}
  \caption{Dynamical-to-stellar mass ratio $\Gamma$ vs. total dynamical mass. Grey points represent GCs and UCDs, where mass estimates are based on integrated velocity dispersions assuming mass traces light from \citet{mieske13} and references therin (with the exceptions of UCD3, UCD 320, UCD 330 and M59-UCD3; \citet{frank11,afanasiev18,voggel18}). Here, stars represent the seven UCDs and two GCs for which AO-assisted data has or will be analyzed. The colored stars represent updated stellar mass measurements after accounting for the central massive BH. Arrows illustrate the effect of the BH. The open grey star represents the M59-UCD3 estimate from \citet{liu15}. In the case of UCD 320 and UCD 330, \citet{voggel18} found their initial dynamical-to-stellar mass ratios were overestimated and did not detect a central massive BH in either object.}
  \label{fig:allucds}
\end{figure}

We can compare our best-fit models with a BH to those without to look at how large a change is caused in the stellar $M/L$.  From our JAM models, the best-fit zero mass BH has a $\Gamma$ of 0.74 ($M/L_{V,dyn} = 3.11$).  This represents our equivalent to the inferred mass from integrated dispersions used to determine the ratio of dynamical to population values \citep[e.g.][]{mieske13}. Therefore, inclusion of a BH in the model reduces the $M/L$ by $\sim$12\%. By contrast, the $M/L$s drop by $>40$\% in all of the other UCDs with central massive BH detections \citep{seth14,ahn17}.    We show the effect on $\Gamma$ of including the BH in M59-UCD3 in Fig.~\ref{fig:allucds}, which, by our determination, does {\em not} have an elevated global dynamical $M/L$ even if there is no BH present.  We note that \citet{liu15} estimated $M/L_{V,dyn}$ to be 4.9$\pm$0.5 based on an integrated dispersion of 77.8$\pm$1.6~km~s$^{-1}$ and $r_{eff} = 25$~pc, shown as an open star in Figure~\ref{fig:allucds}; this estimate is significantly higher than our estimates with or without a BH. \citet{voggel18} showed that measurements of the dynamical-to-stellar mass ratio are easily overestimated using lower spatial resolution data, which demonstrates the importance of high spatial resolution integral field unit (IFU) data for determining the dynamical $M/L$. This is likely the reason for the discrapancy between $M/L$s in \citet{liu15} and our derived value.

In M60-UCD1, VUCD3, and M59cO, the central BH constituted 10--20\% of the total mass of the system \citep{seth14,ahn17}. M59-UCD3 is quite different, with a central BH of only $\sim$2\% of the total mass.  A small, $\sim$4\% mass BH has also been found in Fornax UCD3 \citep{afanasiev18}, but due to much lower $S/N$ data (with similar spatial resolution), this BH was only detectable after assuming isotropy and fixing the stellar $M/L$ to expected stellar population values.  

  In the context of the other UCDs with high BH mass fractions, it is not surprising that the BH mass in M59-UCD3 is less well determined.  Using the conventional definition of the BH sphere of influence ($r_{\rm infl} = GM_{BH}/\sigma^2$) we find $r_{infl} = 0.03-0.07 \asec$, assuming $\sigma = 65.7$~km~s$^{-1}$, for M59-UCD3. Here, the BH sphere of influence range is calculated assuming the best-fit $M_{BH}$ for the axisymmetric Schwarzschild models and JAM models, respectively. Therefore, the BH sphere of influence for M59-UCD3 is approximately an order of magnitude less than what was found for VUCD3, M59cO, and M60-UCD1 \citep{seth14,ahn17}. Furthermore, since our adaptive optics PSF has a core FWHM of 0.165$\asec$ (and the diffraction limit is $0.07\asec$), this likely explains why we can't constrain the lower mass limit of the central BH with all three dynamical models. However, the factor of $\sim$10 range in BH masses in UCDs with similar stellar mass is not particularly surprising given the comparable range of BH masses seen in lower mass (especially spiral) galaxies at a fixed dispersion or stellar mass \citep[e.g.][]{kormendy13,greene16,nguyen17}.

%Therefore, the central massive BH constitutes $\sim$2\% of the total dynamical mass. Furthermore, the addition of the BH reduces the dynamical-to-stellar mass ratio to values comparable to globular clusters. In this case, $\Gamma$ was estimated to be $1.44 \pm 0.15$ prior to the introduction of a central massive BH \citep{liu15}. Our best-fit $\Gamma = 0.67 \pm 0.10$ (3$\sigma$ uncertainties) with the introduction of the SMBH. Figure~\ref{fig:allucds} illustrates this effect for all of the UCDs with BH detections through dynamical modeling. Here, the grey points represent GCs and UCDs. The stars represent objects for which AO-assisted data has or will be analyzed to test for central massive BHs. The red stars indicate objects which have been analyzed. Four UCDs have been found to host central massive BHs: M60-UCD1, VUCD3, M59cO, and Fornax UCD3 \citep{seth14,ahn17}{\bf afanasiev}. A search for central BHs in UCD 320 and UCD 330 yielded a non-detection. However, {\bf voggel18} showed that previous estimates of the dynamical-to-stellar mass ratio were overestimated. These new results suggest these two UCDs were not viable candidates for a SMBH study based on the elevated mass ratios alone. The blue star shows the effect of the SMBH found in M59-UCD3, which we have presented in this paper. 

\subsection{Progenitor Galaxy Properties}

The combination of M59-UCD3's high luminosity and apparent BH strongly suggests it is a tidally stripped remnant of a once more massive galaxy.  We can try to estimate the progenitor mass in several ways.  Assuming that UCDs follow the same $M_{BH}$ vs. Bulge Luminosity relation as galaxies \citep[e.g.][]{mieske13,mcconnell13}, we obtain a progenitor bulge mass of $\sim10^{9}$ $M_\odot$.  This corresponds to the total galaxy mass if we assume an early-type galaxy; in this case, the total galaxy mass would only be $\sim$3$\times$ that of the current UCD, which would make the galaxy remarkably compact, similar to M32's current size and mass.  We can also estimate the mass of the NSC by assuming it is the central two S\'ersic components of the UCD (given that a King + S\'ersic model provides an equally good fit)\citep{pfeffer13}; this component has a mass of $M_{NSC} \equiv 28 \times 10^7$ $M_\odot$ and an effective radius of $r_{eff} = 21$ pc ($0.26\asec$).  Galaxies with similar NSCs in the \citet{georgiev16} sample have stellar masses between 4$\times$10$^9$ and 10$^{11}$~M$_\odot$.  A similar total stellar mass range is found for galaxies with BHs between 10$^6$ and 10$^7$~M$_\odot$ in \citet{reines15}, while the $M-\sigma$ relation implies a galaxy dispersion of $\sim$100~km~s$^{-1}$ \citep[e.g.][]{kormendy13}.  Overall, it appears likely that the original mass of M59-UCD3's progenitor was of order 10$^{10}$~M$_\odot$.

M59 has two UCDs whose progenitor galaxies masses have been estimated to be $\sim 10^9 - 10^{10}$ $M_\odot$. Therefore, we would expect the stars of these disrupted galaxies to be deposited in the outer halo of M59. Using the S\'ersic fits from \citet{kormendy09}, we calculated the total luminosity outside the projected radius of M59-UCD3 and M59cO (both $\sim$10 kpc), which was found to be $L (> 127 \asec) \sim 1.2 \times 10^{10}$ $L_\odot$. Here we have assumed a constant axis ratio and position angle of 0.7 and 164$^\circ$, respectively. We also note that, \citet{liu15} reported a plume with similar luminosity to M59-UCD3 itself, which may be a tidal feature associated with it.  Unless M59-UCD3 was remarkably compact before stripping, it appears likely that this plume, if associated with M59-UCD3, represents just a small fraction of the total amount of mass stripped off the galaxy.  

%  Their study estimated the plume to be a lower limit on the amount of material stripped from M59-UCD3's progenitor.  

%  Although we've shown that there is likely enough material beyond the projected radius of M59-UCD3 and M59cO to account for their estimated progenitor mass, based on $M_{BH}$, it seems likely that M59-UCD3 still contains a large fraction of its progenitor's total stellar mass. }

%If the M59-UCD3 progenitor galaxy follows the same scaling relations between nuclear star clusters (NSCs) and their present day mass \citep[e.g.][]{kormendy13,scott13,vandenbosch16,georgiev16},
%and that UCDs contain $\sim$1\% of their progenitor luminosity \citep[e.g.][]{mieske13}, we estimate the progenitor mass to be $\sim 10^{10} M_\odot$. Furthermore, assuming the inner S\'ersic component is representative of the NSC component of the progenitor \citep{pfeffer13}, the $g-band$ magnitude is $\sim$17.3 and mass $M_{NSC} \equiv 17 \times 10^6 M_\odot$.  
%which is comparable to the nuclei observed in the Virgo Cluster galaxies \citep{cote06}.

\subsection{Further Evidence \& Future Prospects for SMBHs in UCDs}
In this paper, we have also presented deep radio observations of three UCDs around M59 and M60, which showed a detection in only one case (around M59cO).  The radio upper limits in M59-UCD3 and M60-UCD1 are lower than expected for AGN emission based on X-ray detections in these systems, although given the scatter in the fundamental plane, these measurements do not necessarily argue against an SMBH origin for the X-ray emission.  

M59-UCD3 represents the fifth UCD known to host a SMBH. All UCDs with measured BH masses have been shown to have near solar metallicities and $\alpha$-enhancement in the range of [Mg/Fe]$\sim$ 0.1-0.5 \citep{chilingarianmamon08,francis12,sandoval15,janz16}. Both UCDs consistent with no central BH have been shown to have sub-solar metallicities and solar [$\alpha$/Fe] ratios. This could indicate that solar--supersolar metallicity UCDs with alpha enhancement may serve as a secondary indicator of a central massive BH.  

With current telescopes and instrumentation our dynamical modeling effort has, thus far, been limited to either the brightest UCDs at the distance of Virgo/Fornax, or some fainter UCDs at the distance of nearby large galaxies such as Centaurus A. This is due to a combination of the need to resolve the BH sphere of influence, as well as limits on the faintest sources observable with adaptive optics corrections. As the next generation telescopes come online, such as the Extremely Large Telescope (ELT), we will be able to significantly increase the number of UCDs for which we have the capability to run dynamical models. For example, the new High Angular Resolution Monolithic Optical and Near-infrared Integral Field Spectrograph (HARMONI), which is to be mounted on ELT has an estimated adaptive optics corrected angular resolution of $0.01 - 0.03 \asec$ \citep{cunningham08}. Assuming two hypothetical BHs with $M_{BH} = 10^7$ $M_\odot$ and $M_{BH} = 10^6$ $M_\odot$ and corresponding integrated dispersions of $\sim$50~km~s$^{-1}$ and $\sim$30~km~s$^{-1}$, the BH sphere of influence would be $\sim$16 pc and $\sim$5 pc, respectively. If we require the BH sphere of influence to be at least 0.05$\asec$ in size for us to resolve it with ELT/HARMONI, then we could theoretically resolve these BHs out to 19 Mpc (the distance of Fornax/Virgo) for the $10^6$ $M_\odot$ BH and an astonishing 66 Mpc (the distance of the Perseus cluster and many others) for the $10^7$ $M_\odot$ BH.

{\it Facilities:} \facility{Gemini:Gillett (NIFS/ALTAIR)},
\facility{HST (WFC3)}, \facility{The National Radio Astronomy Observatory is a facility of the National Science Foundation operated under cooperative agreement by Associated Universities, Inc.}, \facility{These results are partially based on data obtained from the Chandra Data Archive}

{\em Acknowledgments: C.P.A.~and A.C.S.~acknowledge financial support from HST grant GO-14067 and NSF AST-1350389. MC acknowledges support from a Royal Society University Research Fellowship. JS acknowledges support from NSF grant AST-1514763 and the Packard Foundation. AJR was supported by National Science Foundation grant AST-1515084 and as a Research Corporation for Science Advancement Cottrell Scholar. RMcD is the recipient of an Australian Research Council Future Fellowship (project number FT150100333).} 
%\bibliography{archive}
%\bibliographystyle{apj}
\newcommand{\noop}[1]{}

\end{document}